\documentclass[11pt]{extarticle}
\usepackage{amsfonts,latexsym,amssymb}

\usepackage{epsfig}

\usepackage[applemac]{inputenc}
\usepackage{amsfonts}

\advance\voffset by -15mm
\advance\hoffset by -20mm
\usepackage{epsfig}
\usepackage[applemac]{inputenc}
\usepackage{amsfonts}
\usepackage{amsmath}
 \usepackage{psfrag}
  \usepackage{rotating}

\def \N{{\mathbb  N}}

\def \R{{\mathbb R}}
\def\E{{\mathbb E}}

\def\ep{\varepsilon}
\renewcommand{\le}{\leqslant}
\renewcommand{\ge}{\geqslant}
\renewcommand{\hat}{\widehat}
\renewcommand{\phi}{\varphi}

\numberwithin{equation}{section}

\newcommand{\eps}{\varepsilon}

\newtheorem{lemma}{Lemma}
\newtheorem{theorem}{Theorem}

\newtheorem{remark}{Remark}
\newtheorem{example}{Example}
\newtheorem{proposition}{Proposition}

\begin{document}
\title{\bf  Stochastic models and numerical algorithms\\
            for a class of regulatory  gene networks 
    }
\author{ Thomas Fournier,  Jean-Pierre Gabriel, Christian Mazza, Jer\^ome Pasquier\thanks{Department of Mathematics, University of Fribourg,Chemin du Mus\'ee 23,CH-1700 Fribourg, Switzerland, christian Mazza@unifr.ch}\\
 Jos\'e Galbete and Nicolas Mermod\thanks{Institute of Biotechnology, University of Lausanne, CH-1015 Lausanne, Switzerland, nicolas.mermod@unil.ch}}

\date{ February 2008 }

\maketitle

\thispagestyle{empty}

\noindent{\bf Regulatory gene networks contain generic modules, like those
involving feedback loops, which are essential for
the  regulation
of many biological functions  \cite{Guido}.
  We consider a class of self-regulated genes which are the
  building blocks of many regulatory gene networks, and
  study the steady-state distribution of the associated Gillespie algorithm
  by providing efficient numerical algorithms. 
 We also study a regulatory gene 
  network of interest in gene therapy, using mean-field models with time delays.
  Convergence of the related time-nonhomogenous Markov chain is established
  for a class of linear catalytic networks with feedback loops.
 
 }
 
\medskip


\medskip

\section{Introduction\label{s.intro}}
Modeling of the regulation of all genes in a given cell is a tantalizing problem in biology and medicine (see, e.g. \cite{Guido}). Recent developments allow rapid experimental determination of the expression of nearly all genes in a given biological setting, to an extent that in depth analysis and proper mathematical understanding of these vast arrays of data has become limiting. Qualitative models of regulatory networks, where particular genes code for proteins that activate or repress other genes, are being assembled, but models taking the stochastic and quantitative nature of gene regulation remain scarce, and they often rely on assumptions or simplifications that rest untested experimentally. Thus, it would be useful to build validated mathematical models of particular regulatory modules, as a first step towards constructing models of genome-wide gene expression. 

\begin{figure}
  \centering
\psfrag{A}{{\sc Promoter is OFF}}
\psfrag{B}{{\sc Promoter is ON}}
\psfrag{2}{{\bf OFF}}
\psfrag{1}{{\bf ON}}
\psfrag{4}{Expressed gene}
\psfrag{5}{Silenced gene}
\psfrag{6}{\tiny Intermediate reaction}
\psfrag{7}{\tiny Protein}
\psfrag{9}{\tiny production}
\psfrag{8}{\tiny Protein degradation}
\includegraphics[width=\textwidth]{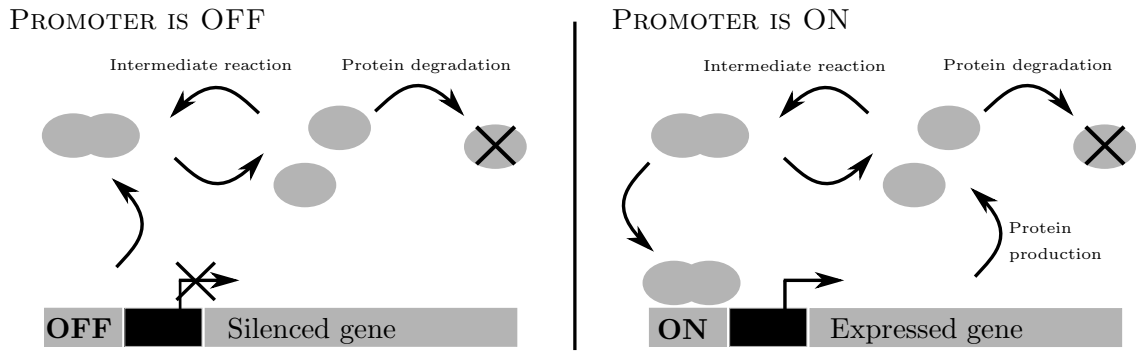}
  \caption{The self-regulated gene}
  \label{fig0}
\end{figure}

Here, we consider a class of self-regulated genes, as depicted in Figure \ref{fig0}. This auto-regulated module is a very common building block of many gene networks, as it may form the basis of stochastic gene switches that contribute to biological decisions such as cell differentiation, and has been studied extensively in the literature in some special settings, as in \cite{Peccoud}, \cite{Wolynes} or \cite{Kepler}.  In a previous work, \cite{Fournier}, we provided the exact steady-state distribution of the stochastic expression level of the autoregulated gene using the Gillespie algorithm in a general setting. We will present a direct version of the method and study more deeply this stationary distribution by providing efficient numerical algorithms. We will also consider a synthetic regulatory network acting as a genetic switch that was studied in living cells \cite{Mermod}. 
\medskip

 \medskip

\noindent {\bf A self-regulated gene}
\medskip

The system is composed of a promoter and a gene, as schematized in Figure \ref{fig0}.  As stated previously, one source of molecular noise is the random nature of the states taken by the promoter (on/off).   Figure \ref{fig0} shows protein monomers produced by the RNA polymerase during the transcription and translation processes. Protein monomers react quickly to form dimers: we assume a quasi-equilibrium where fast reactions equilibrate instantaneously. 
For a global amount of n proteins, the proportion of  dimers at quasi-equilibrium is a well defined function of n. 
Dimers can bind to some sites near the promoter, and therefore enhance  transcription, corresponding to a positive  feedback loop. These binding events can be assumed to be fast with respect to events like protein formation. They are however included in some chain of events which ends with a state where  the right positioning of the RNA polymerase is possible. This will correspond to the on state ${\cal O}_1$. When these conditions are not satisfied, the promoter is off ${\cal O}_0$. 
The rates of transitions between these two states are functions of the proportion of dimers, and therefore of n when the cell contains n proteins.  These random events are usually modeled by supposing that the probability that the promoter switches from the off to on state in a small time interval of length $h\approx 0$  is of order g(n)h for n   proteins, where the function g can be chosen according to the specificity of the setting.  To be as general as possible, and to eventually allow negative feedback loops, we also assume that the probability of transition of the reverse reaction is given by some function $\kappa(n)$. Basal activity is introduced by supposing that g(0) is positive, so that the required conditions for an eventual transcription event can be realized without protein dimers. 
The remaining involved chemical reactions are essentially protein monomers production and degradation, which are summarized in Figure \ref{fig0}. Transcription is stopped when the promoter is off, so that we assume that the probability $\mu_0 h$ that a protein is created during a small time interval of length h vanishes, with $\mu_0=0$. When the promoter is on, transcription is possible, and the probability that a transcription event occurs is of order $\mu_1 h$.  Degradation of protein dimers is summarized by the rate $\nu(n)$, for some function $\nu$, which is usually linear as a function of $n$.  
The time evolution of the state of this self regulated gene is described by a pair of time continuous stochastic process N(t) and Y(t), where N(t) gives the number of proteins present in the cell at time t and where Y(t) takes the values 0 and 1 corresponding to the off and on states of the promoter. The usual way of simulating N(t) and Y(t) proceeds by running the Gillespie algorithm (see e.g. \cite{GillespieB}, and \cite{Cao}). The mean steady state expression level is thus obtained through Monte-Carlo experiments.

\medskip

\noindent {\bf A regulatory network for efficient control of transgene expression}
\medskip

\begin{figure}
  \centering
  \includegraphics[width=0.7\textwidth]{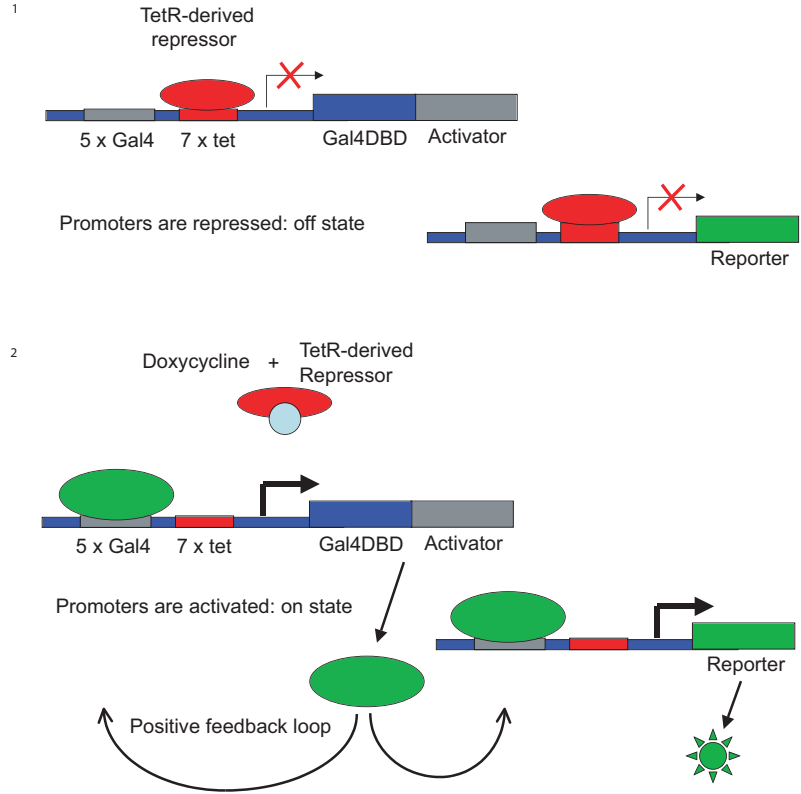}
  \caption{The regulatory network}
  \label{fig1}
\end{figure}

\noindent A more elaborate gene network consists of three genes. A first gene encodes a transcriptional repressor. Because this gene is expressed from an unregulated promoter, it mediates a stable number of repressor. This repressor binds to and inhibits the promoters of the two other genes, coding for a transactivator protein and for a quantifiable or a therapeutic protein, respectively (Figure \ref{fig1}A). The activity of the repressor is inhibited by doxycycline, a small antibiotic molecule that acts as a ligand of the repressor and thereby controls its activity. Addition of the antibiotic will inhibit the repressor and relieve repression, allowing low levels of expression of the regulated genes and synthesis of some transactivator protein. This, in turn, allows further activation of the two regulated genes, in a positive feedback loop (Figure \ref{fig1}B). When introduced in mammalian cells, this behaves as a signal amplifier and as a potent genetic switch, where the expression of a therapeutic gene can be controlled to vary from almost undetectable to very high levels in response to the addition of the antibiotic to the cells (\cite{Mermod}, and \cite{Fournier}). 
\bigskip

\noindent {\bf Results}
\bigskip

\noindent Section \ref{s.transcription.quasi} considers the Gillespie algorithm for simulating
the time evolution of the number of proteins $N(t)$ and of the state of of the related
promoter $Y(t)$, by focusing on the associated steady state distribution $\pi$. In a previous work \cite{Fournier}, we gave
an explicit formula for the steady-state associated to self-regulated genes, based on the embedded jump chain of the time continuous Markov process. Here we introduce a direct version dealing with the time continuous process. For concrete computation, we have to use a bounded state space with a total number of proteins that can not exceed a fixed but arbitrary integer $\Lambda$. ALGORITHM I gives an efficient way of computing $\pi$ and a useful tightness argument to show that the sequence of steady-state distributions measures indexed by $\Lambda$ converges as $\Lambda\to\infty$ to the unique invariant distribution of the process defined on the unbounded state space $\N\times\{0,1\}$. We also provide information on the variance of the gene
product at steady state using generating functions and differential equations.

Section
\ref{s.meanfield} proposes a mean field model with time delays, generalizing  a model considered
recently in this setting by \cite{MeanField},
by including stochastic signals related to promoters. The feedback rates
$g(N(t))$ and $\kappa(N(t))$ are replaced by
$g(\E(N(t-\theta)))$ and $\kappa(\E(N(t-\theta)))$. In \cite{Fournier}, we studied the
 regulatory gene network
  in living cells; the obtained experimental results were in good
 concordance with the model's predictions. 
The related functions $E(t)=\E(N(t))$
and $G(t)=P(Y(t)=1)$  sastisfy the time delayed differential system
$$
  \frac{dE}{dt}=\mu G(t)-\nu E(t),\ \ \ \frac{dG}{dt}=g(E(t-\theta))(1-G(t))-\kappa(E(t-\theta))G(t),
$$
which can be deduced from the chemical master equation, see Section \ref{network}.
As it is well known, this kind of differential systems can possess oscillating or periodic solutions,
see e.g. \cite{Delayed}.
We show that there is a globally asymptotically stable equilibrium point when $g$ is such that
$g(n)/n$ is decreasing as function of $n$ and $\kappa(n)\equiv \kappa$. We next provide ALGORITHM
 II
for computing the steady state variance. Section \ref{s.two} deals with
two time scales stochastic simulations and processes evolving at quasi-equilibrium. We also consider
a generic  dimerization process which occurs in most biochemical reaction networks, and provide an efficient
ALGORITHM III for computing the first two moments of the related steady state distribution,
which are then used when dealing with systems evolving at quasi-equilibrium.
Finally,
Section \ref{network} focus on the regulatory network; we model extrinsic and intrinsic noise using
a mean-field model, which permits to study the fluctuations of the variance of the number
of therapeutic proteins as function of the number of doxycycline molecules. 


\section{Mathematical models related to  the self-regulated gene\label{s.transcription.quasi}}

In what follows, we consider  the time continuous Markov chain known as {\it the Gillespie algorithm} for 
simulating the self regulated gene with arbitraries feedback
mechanisms, and give precise formulas for the related steady state distribution $\pi$.
We shall see in Section \ref{s.meanfieldnet} a semi-stochastic or mean field
model for the therapeutic network. The related steady state distribution
is obtained as the product of steady state distributions of sub modules corresponding
to self regulated genes. A complete understanding of basic modules like the self regulated
gene is thus fundamental for understanding the global network, see e.g. \cite{Guido} and
Section \ref{s.meanfieldnet}. For more details on the model, see \cite{Fournier}.

The module is composed of a promoter and a gene. Its time evolution is given by
the following  set of chemical reactions:
$$
{\cal P}{\overset{\nu (n) }{\longrightarrow}}\emptyset,\ \ \emptyset{\overset{ \mu_l }{\longrightarrow}}{\cal P},\ l=0,1,
$$
represents degradation of gene product when they are $n$
molecules,
 here proteins (${\cal P}$), and protein production,
where $l=0$ means
that the promoter is off: no transcription factor (a complex composed of gene product) is bound to some operator sites
near the promoter, so that
the RNA polymerase can't bind well in the neighborhood of
the basal promoter. We assume here that
the transcription rate $\mu_0$ is such that
$\mu_0\approx 0$.
When $l=1$, meaning that
the promoter is on, transcription occurs at a rate $\mu_1=\mu$. The fluctuations of the state
of the promoter are described by the following reaction
$$
{\cal O}_0 \underset{\kappa(n) }{\overset{ g(n)}{\longleftrightarrow}}{\cal O}_1,
$$
where ${\cal O}_l$, $l\in\{0,1\}$, indicates the state of the promoter. The transitions from the on to off states
${\cal O}_1 {\overset{\kappa(n)}{\longrightarrow}}{\cal O}_0$
occur at rate $\kappa(n)$, and the reverse reactions at rate $ g(n)$. Here
$ g$ and $\kappa$ are two functions of the number of proteins modeling
positive and negative feedback loops.

 Basal activity is introduced at the level of the reaction
${\cal O}_0 {\overset{ g(n)}{\longrightarrow}}{\cal O}_1$, by
supposing that $ g>0$.
The Gillespie algorithm
for simulating the above chemical reactions is
a bivariate Markov process
$\eta(t)$
 $=( N(t), Y(t))$, with $ N(t)\le \Lambda$ and
  $ Y(t)=0,1$,
 where $ N (t)$ denotes
the number of proteins  
 present in the cell at  time $t$, and
 $ Y(t)$ represents the state of the operator. 
 The time evolution of $ Y(t)$ is coupled to that of $ N(t)$
to model auto-regulation, using  functions
$ g$ and $\kappa$. For  small time interval
$(t,t+h)$, $h\approx 0$, 
the probability that the operator switches from the off to
the on state is of order $ g( N(t))h$.

 Let $p_n^0(t)=P(N(t)=n,Y(t)=0)$ and $p_n^1(t)=P(N(t)=n,Y(t)=1)$ give the probability of having $n$ proteins
at time $t$
when the states of the promoter are ${\mathcal O}_0$ and ${\mathcal O}_1$, respectively. 
We assume here that $0\le n\le \Lambda$ for some fixed but arbitrary integer $\Lambda$.
The related Gillespie algorithm is given as a time-continuous Markov chain
$\eta(t)=(N(t),Y(t))$ (see e.g. \cite{Gillespie}),
  where
 $N(t)\in\{0,1,\cdots,\Lambda\}$ and $Y(t)\in \{0,1\}$, with
 transition rates given by
 $$P((n,y),(n+1,y))=\mu_y,\ P((n,y),(n-1,y))=\nu(n),$$
 $$P((n,y),(n,1-y))=\kappa(n)\hbox{ when }y=1,\hbox{ and },$$
 $$P((n,y),(n,1-y))=g(n)\hbox{ when }y=0.$$
The  {\it chemical master equation} associated to the reaction scheme 
is then given by
\begin{equation}\label{Master}
\frac{{\rm d}p_n^s(t)}{{\rm d}t}=\mu_s(p_{n-1}^s(t)-p_n^s(t))+\nu(n+1)p_{n+1}^s(t)-\nu(n)p_n^s(t)
\end{equation}
$$+(-1)^s(\kappa(n)p_n^1(t)-g(n)p_n^0(t)),$$
where $s\in\{0,1\}$,
see e.g. \cite{Kepler}. 

\begin{figure}\label{fig:strip}
\psfrag{m}{$\mu$}
\psfrag{v}{$\nu (n)$}
\psfrag{k}{$\kappa (n)$}
\psfrag{g}[l]{$g(n)$}
\psfrag{a}{$(n-1,1)$}
\psfrag{b}{$(n,1)$}
\psfrag{c}{$(n+1,1)$}
\psfrag{d}{$(n-1,0)$}
\psfrag{e}{$(n,0)$}
\psfrag{f}{$(n+1,0)$}
\includegraphics[width=\textwidth]{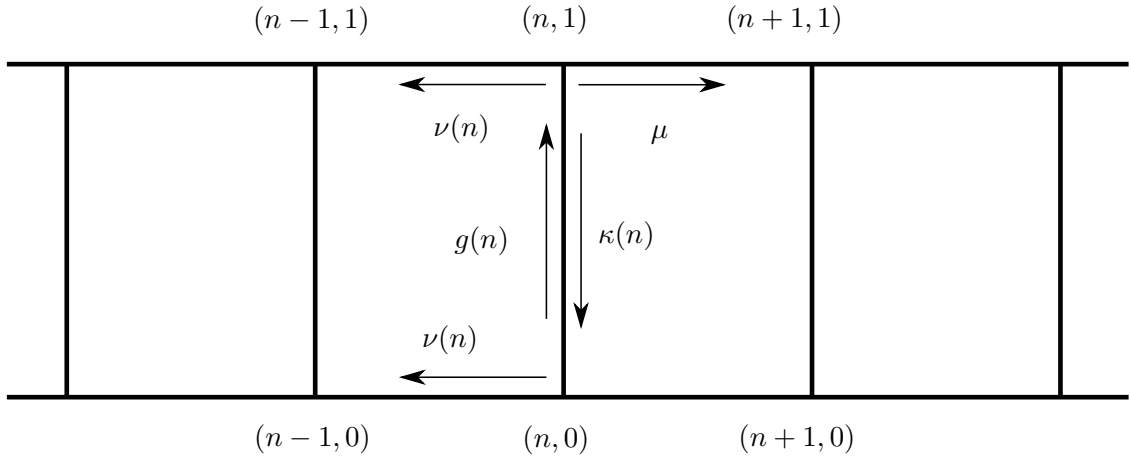}
\caption{Visualization of the state space as a strip. The possibles transitions are represented by the arrows with corresponding rates.}
\end{figure}

The steady-state distribution $\pi$ associated with (\ref{Master}) is obtained by letting
$t\to\infty$: $\pi$ is defined as
$$\pi_n(0)=\lim_{t\to\infty}p_n^0(t)\hbox{ and }\pi_n(1)=\lim_{t\to\infty}p_n^1(t),$$
and solves the linear system obtained from (\ref{Master}) by imposing
${\rm d}p_n^s/{\rm d}t=0$:
$$
0=\mu_s(\pi_{n-1}(s)-\pi_n(s))+\nu(n+1)\pi_{n+1}(s)-\nu(n)\pi_n(s)$$
$$
   +(-1)^s(\kappa(n)\pi_n(1)-g(n)\pi_n(0)),\ s=0,\ 1.$$
   $\pi_n(0)$ is
the probability to find $n$ proteins and that
          the
promoter is off; $\pi_n(1)$ is defined similarly but for the
on
         state. The probability of observing $n$ proteins at equilibrium
         is just $\pi_n(0)+\pi_n(1)$. 
         
 In what follows, we derive the steady state distribution $\pi$. This probability
 measure is used in the mean-field delayed model of Section \ref{s.meanfield} and
 in Section \ref{network} for the study of the network.


\subsection{Computing the steady-state\label{SteadyCompute}}



In this Section, we assume that $\mu_0=0$, and that
$\mu_1=\mu$, where an explicit formula for the steady-state
is available. The degradation propensity function $\nu(n)$ and the feedbacks $\kappa(n)$ and $g(n)$ are arbitrary positive functions.

The method of generating functions can be used in some particuliar special cases to compute the invariant measure, see \cite{Peccoud} for the simple case whitout feedback with $\nu(n)=\nu\cdot n$, $\kappa(n)\equiv\kappa$ and $g(n)\equiv g$, or \cite{Wolynes} for the case with linear negative feedback $\nu(n)=\nu\cdot n$, $\kappa(n)\equiv\kappa\cdot n$ and $g(n)\equiv g$. Although this method provides a powerful tool for analytic description, the method of generating functions is very particular in the sense that a little change in the form of one of the feedback propensity function can induce major changes in the generating function, and for each particular propensity function one has to derive the whole set of equations anew. Furthermore, an explicit form for the generating function can only be found when the feedback propensity functions are simple, either constant or linear in the protein numbers. In practice, the feedback propensity functions are related to the number of sites in the promoter on which the proteins bind, either directly in monomer form or in more complicated bound forms like dimers or higher order polymers, see \cite{Dill}.

In \cite{Fournier}, we presented a general formula for the steady-state for arbitrary degradation and feedback propensity functions. The method relies on the asymptotic behaviour of the jump matrix of the embedded discrete jump chain. We provide here a direct version of this method allowing to compute the invariant distribution of the time continuous Markov process directly. We recall that we consider a bounded total number of protein $N(t)\leq\Lambda$, a restriction that is biologically meaningful due to the finite volume of a cell but that is mainly supposed for technical reasons since the formula is recursive and we have to find a starting point $(\pi_\Lambda(0),\pi_\Lambda(1))$ to begin with. However, the condition is not restrictive and we show in Theorem \ref{thm:ConvLambda} that the sequence of steady-state distributions indexed by the boundary $\Lambda$ converges weakly to the unique invariant distribution of the unbounded process on $\N\times\{0,1\}$.

Let us define the transfer matrices
$$\alpha_{n}=\frac{\nu(n+1)}{\mu}\left [\begin{array}{cc}\frac{\kappa(n)+\mu}{g(n)+\nu(n)}&1\\\frac{\kappa(n)}{g(n)+\nu(n)}&1\end{array}\right ],\ 0<n<\Lambda,\quad\alpha_{0}=\frac{\nu(1)}{\mu}\left [\begin{array}{cc}\frac{\kappa(0)+\mu}{g(0)}&1\\\frac{\kappa(0)}{g(0)}&1\end{array}\right ],$$
and the vector $w_\Lambda=(\kappa(\Lambda),g(\Lambda)+\nu(\Lambda))$.

\begin{theorem}\label{Steady}
For $0\leq n\leq \Lambda$, the invariant distribution $$\pi_n=(\pi_n(0),\pi_n(1))=\lim_{t\to\infty}\left(P((N(t),Y(t))=(n,0)),P((N(t),Y(t))=(n,1))\right)$$ 
 of the time-continuous Markov process 
$\{\eta(t)\}_{t>0}$ on the strip $\{0,1,\dots, \Lambda\}\times\{0,1\}$ is given by 
$$
\pi_n=\frac{w_\Lambda \alpha_{\Lambda-1}\alpha_{\Lambda-2}\cdots\alpha_{n}}{Z_\Lambda}\;\text{ for }\; 0\leq n<\Lambda,\;\text{ and }\;\pi_\Lambda=\frac{w_\Lambda}{Z_\Lambda},
$$
with the normalization constant $$\displaystyle Z_\Lambda=w_\Lambda\cdot (1,1)^T+\sum_{j=0}^{\Lambda-1}w_\Lambda\alpha_{\Lambda-1}\alpha_{\Lambda-2}\cdots\alpha_{j}\cdot(1,1)^T.$$
\end{theorem}

\noindent{\bf Proof :} At equilibrium, equations \ref{Master} reads
\begin{align*}
0=&\pi_0 R_0+\pi_1D_1,\\
0=&\pi_{n-1}U+\pi_nR_n+\pi_{n+1}D_{n+1},\quad 0<n<\Lambda,\\
0=&\pi_{\Lambda-1}U+\pi_{\Lambda}R_\Lambda.
\end{align*}
with the $2\times 2$ matrices $U=\left[\begin{array}{cc}0&0\\0&\mu\end{array}\right ],$ $D_n=\left[\begin{array}{cc}\nu(n)&0\\0&\nu(n)\end{array}\right ]$
 and $$R_n=\left[\begin{array}{cc}-(g(n)+\nu(n))&g(n)\\ \kappa(n)&-(\kappa(n)+\nu(n)+\mu)\end{array}\right ]$$ for $1\leq n\leq \Lambda-1, $
and the boundaries
$R_0=\left[\begin{array}{cc}-g(0)&g(0)\\\kappa(0)&-(\kappa(0)+\mu)\end{array}\right ],$ $$ D_\Lambda=\left[\begin{array}{cc}\nu(\Lambda)&0\\0&\nu(\Lambda)\end{array}\right ] \text{  and } R_\Lambda=\left[\begin{array}{cc}-(g(\Lambda)+\nu(\Lambda))&g(\Lambda)\\\kappa(\Lambda)&-(\kappa(\Lambda)+\nu(\Lambda))\end{array}\right ].$$
Simple linear algebra shows that the above defined matrices $\alpha_n$, $0\leq n<\Lambda$, satisfy the relation $\pi_n=\pi_{n+1}\alpha_n$. Indeed one only has to check that for $0<n<\Lambda$, the $2\times 2$ matrices $\alpha_{n-1}U+R_n$ are invertible and that the matrices $\alpha_n$ solve the matrix continuous fraction
\begin{align*}
\alpha_0=&-D_1R_0^{-1},\\
\alpha_n=&-D_{n+1}(\alpha_{n-1}U+R_n)^{-1},\quad 0<n<\Lambda.
\end{align*}
\hfill$\square$\medskip\\

The formula given in Theorem \ref{Steady} must be used with care  numerically since, when $\Lambda$ is large, both the numerator and denominator rapidly diverge. It can be improved with the following normalization algorithm, that is exactly the same as the one used for the embedded jump chain in \cite{Fournier}. Let $\R_{\geq 0}^{2}:=\{w=(w_1,w_2),w_1,w_2\in\R_{\geq 0}\}$ with the $1$-norm $\|w\|:=w\cdot  (1,1)^T$.
  
 \bigskip

  \centerline{\bf ALGORITHM I}
  \bigskip
  
  \indent {\bf (STEP 1):} Define $\tilde v_n$ for $n=\Lambda-1$ to $0$ as
  $$\tilde v_\Lambda:=\frac{w_\Lambda}{\| w_\Lambda\|},\quad\text{and}\quad \tilde v_n:=\frac{\tilde v_{n+1}\alpha_n}{\|\tilde v_{n+1}\alpha_n\|}.$$
  \indent {\bf (STEP 2):} Given the  $\tilde v_n$, define $ v_0=\tilde v_0$ and, for $n=1$ to $\Lambda$, set
  $$v_n:=\frac{\tilde v_n}{\|\tilde v_n \alpha_{n-1}\|\cdot\|\tilde v_{n-1} \alpha_{n-2}\|\cdots\|\tilde v_1 \alpha_{0}\|}.$$
  \indent {\bf (STEP 3):} Compute the steady-state distribution as
  $$\pi_n=\frac{v_n}{V_\Lambda},\qquad\text{ where }\quad V_\Lambda:=\sum_{i=0}^{\Lambda}v_i\cdot{\bf 1}.$$
  
\noindent It immediately results from their definition that the $\tilde v_n$ and $v_n$ satisfy $\|\tilde v_n\|=1$, 
$$v_n=\frac{\tilde v_\Lambda\alpha_{\Lambda-1}\alpha_{\Lambda-2}\cdots\alpha_{n}}{\|\tilde v_\Lambda\alpha_{\Lambda-1}\|\cdot\|\tilde v_{\Lambda-1}\alpha_{\Lambda-2}\|\cdots\|\tilde v_{n+1}\alpha_{n}\|\cdot\|\tilde v_{n}\alpha_{n-1}\|\cdots\|\tilde v_1\alpha_{0}\|},$$
the denominator of the above expression is independent of $n$ and $\tilde v_\Lambda$ is proportional to $w_\Lambda$. Hence $v_n$ is proportional to the invariant measure $\pi_n$, and {\bf (STEP 3)} of the algorithm effectively compute the actual steady-state distribution.

Proposition \ref{bounds} below provides conditions under which
   the normalization constant $V_\Lambda$ remains bounded as $\Lambda$ is large. The function $\nu(n)$ gives the monomer degradation rates for $n$ proteins, 
   and is assumed to  be increasing with  $ \nu (0)=0$, and  strictly positive for $n\geq 1$. Usually, $\nu(n)$ is taken to be a constant times $n$, here we assume the less restrictive condition that 
   $\displaystyle\inf\limits_{n\geq 1} \nu(n)/n$ is stricly positive
to allow situations where for example proteins that are present as chemical complexes (dimer, trimer,...) can not be degradated, or situations where $\nu(n)/n\to\infty$ as $n\to \infty$.

 \begin{lemma}\label{Inequality}If $\displaystyle\inf\limits_{n\geq 1} \nu(n)/n$ is stricly positive, there exists a constant  $k>0$ depending only on $\mu,\,\nu,\,\kappa,\, g$ (and not on $\Lambda$) such that for all $n\geq 1$, $\|\tilde v_n\alpha_{n-1}\|\geq n k.$
\end{lemma}
{\bf Proof:} Each $\tilde v_j$ lies in the line segment $S\subset \R^2$ between the points $(0,1)$ and $(1,0)$ and depends on $\Lambda$. To break this dependence, we prove the results for an arbitrary vector $v=(t,1-t)\in S$, $t\in [0,1]$. \\
$$v\alpha_n=\frac{\nu(n)}{\mu}(\star,t+(1-t))=\frac{\nu(n)}{\mu}(\star,1),$$
with $\star>0$ for all $n$. Hence, uniformly in $S$,
$$\frac{\|v\alpha_n\|}{n}\geq \frac{\nu(n)}{n\mu}\geq \inf\limits_{n\geq 1} \frac{\nu(n)}{n\mu}=:k>0.$$
\hfill$\square$\medskip\\
With Lemma \ref{Inequality} we can give bounds uniformly in $\Lambda$:
\begin{proposition}\label{bounds}  Assume that $\displaystyle\inf\limits_{n\geq 1} \nu(n)/n$ is stricly positive. There exists $M>0$, depending only on $\mu,\,\nu,\,\kappa,\, g$ (and not on $\Lambda$), such that
$$\displaystyle 1\leq V_\Lambda=\sum_{n=0}^{\Lambda}\|v_n\|\leq M.$$
\end{proposition}
 {\bf Proof :} Notice that $\|v_0\|=\|\tilde v_0\|=1$. The $1$-norm of $v_n$  is $$\|v_n\|=\frac{\|\tilde v_n\|}{\|\tilde v_n\alpha_{n-1}\|\cdots\|\tilde v_1\alpha_0\|}=\frac{1}{\|\tilde v_n\alpha_{n-1}\|\cdots\|\tilde v_1\alpha_0\|}$$ and with Lemma \ref{Inequality} 
$$V_\Lambda=1+\sum_{n=1}^{\Lambda}\|v_n\|\leq 1+\sum_{n=1}^{\infty}\frac{k^{-n}}{n!}
=e^{1/k}=:M.$$

The first aim of lemma \ref{Inequality} is to show that the preceding algorithm is efficient. But this lemma can also be used to demonstrate that the steady-state distribution of the continuous-time process $\eta(t)$ converges when $\Lambda\to\infty$. Moreover, we can show that this limiting distribution is the invariant distribution of the process on the unbounded strip.
\\
It is necessary to adapt our notations in order to show the dependency in $\Lambda$. Henceforth, we will write $\pi_n^{(\Lambda)}$ and $v_n^{(\Lambda)}$ instead of $\pi_n$ and $v_n$.
\\
Until now, we have considered a finite state-space by fixing a maximum number $\Lambda$ of proteins. It is always easier to deal with finite Markov chains, but the main reason is because our algorithm to compute the invariant measure works in this case. Even if this model is realistic (an organism cannot contain an infinite number of proteins), it is interesting to show that the steady-state does not depend asymptotically on this maximum number $\Lambda$ of proteins. In other words, we want to show that under a sufficient condition, the invariant measure converges in $\Lambda$. We define $\pi^{(\Lambda)}=(\pi_n^{(\Lambda)})_{0\leq n\leq\Lambda}$ and consider the familiy of probalility measures $\Pi=(\pi^{(\Lambda)})_{\Lambda\in\N}$ embedded in $\N\times\{0,1\}$.

\begin{theorem}\label{thm:ConvLambda} The sequence of invariant measures $\pi^{(\Lambda)}$ converges weakly as $\Lambda\to \infty$ to the invariant distribution of the process defined on the unbounded strip. \end{theorem}

\noindent{\bf Proof:} A sequence $\{P_{n}\}$ of probability distributions on a countable and discrete state space  $E$ converges to the probability distribution $P$ on $E$ if and only if each of its subsequences $\{P_{n'}\}$ contains a further subsequence $\{P_{n''}\}$ that converges to $P$. A family $\Pi$ of probability distributions on $E$ is called \emph{relatively compact} if every sequence of elements of $\Pi$ contains a convergent subsequence (to a probability distribution on $E$), and \emph{tight} if for every positive $\epsilon$ there exists a compact set $K$ such that $P(K)>1-\epsilon$ for all $P$ in $\Pi$. \emph{Tightness} implies \emph{relative compactness}, see e.g.~\cite{Billingsley}.\\
We first show that the family of probalility distributions $\Pi=(\pi^{(\Lambda)})_{\Lambda\in\N}$ is tight. Lemma \ref{Inequality} implies that there exists $k$ not depending on $\Lambda$ such that $$\|v_n^{(\Lambda)}\|\leq\frac{1}{n!\,k^n},\quad n\geq 1.$$ Since $V_\Lambda\geq 1$, we have also $$\|\pi_n^{(\Lambda)}\|=\frac{\|v_n^{(\Lambda)}\|}{V_\Lambda}\leq\frac{1}{n!\,k^n},\quad n\geq 1.$$ Hence, for all $\epsilon>0$ there exists $M_\epsilon$ (not depending on $\Lambda$) such that $$\sum_{j=M_\epsilon}^\infty\|\pi_j^{(\Lambda)}\|\leq\sum_{j=M_\epsilon}^\infty\frac{1}{j!\,k^j}<\epsilon.$$ 
Consequently, $\Pi$ is relatively compact and there exists a convergent subsequence $(\pi^{(\Lambda_k)})_{k\in\N}$ of $\Pi$. Define $\pi^{(\infty)}=\lim_{k\to\infty}\pi^{(\Lambda_k)}$. We check now that $\pi^{(\infty)}$ is the invariant distribution of the continuous-time process defined on the unbounded strip. For each $n\in\N$ we have
\begin{align*} \pi_n^{(\infty)} &= \lim_{k\to\infty}\pi_n^{(\Lambda_k)} \\ &= \lim_{k\to\infty}(\pi_{n+1}^{(\Lambda_k)}\,D_{n+1}+\pi_n^{(\Lambda_k)}\,R_n+\pi_{n-1}^{(\Lambda_k)}\,U) \\ &= \pi_{n+1}^{(\infty)}\,D_{n+1}+\pi_n^{(\infty)}\,R_n+\pi_{n-1}^{(\infty)}\,U. \end{align*}
This shows that $\pi^{(\infty)}$ is indeed an invariant distribution of the limit chain. In fact, what preceded is also valid for any converging subsequence. Besides, the invariant distribution is unique because the process is irreducible. Thus, we can conclude that $\pi^{(\Lambda)}$ converges as $\Lambda\to \infty$ to $\pi^{(\infty)}$ which is the invariant distribution of the process defined on the unbounded strip.$\hfill\square$


\subsection{The method of generating functions for the mean and variance\label{s.generating}}
 We consider the problem  of computing the mean and variance of the
gene product $N(t)$ at steady-state, that is when $t$ is large, using
generating functions. As discussed in Section \ref{SteadyCompute}, generating functions allows in some simple cases to compute the steady-state distribution, see e.~g.~\cite{Peccoud} or \cite{Wolynes}, with simple feedback functions. Here we show that even when the feedbacks are arbitrary, the method can be used to gain insight in the relations between variance, mean and probability to be ON. To avoid boundary conditions, we suppose here that the number of protein is arbitrary ($\Lambda=\infty$), and the only asumption concerning the propensity functions is that $\nu(n)=\nu \ n$, i.~e.~degradation is directly proportional to the number of proteins, while $\kappa(n)$ and $g(n)$ are arbitrary positive functions and $\mu_0$ is not necessarily $0$.

Let $\pi_n(y)=\lim\limits_{t\to\infty}P(N(t)=n,Y(t)=y)$ and
$$\alpha(z)=\sum_{n\ge 0}\pi_n(1)z^n,\qquad \qquad
\beta(z)=\sum_{n\ge 0}\pi_n(0)z^n,$$
 be the partial generating functions related to the steady-state, and 
$$R(z):=\sum_{n\ge 0}(\pi_n(1)\kappa(n)-\pi_n(0)g(n))z^n.$$
From the master equation (\ref{Master}) at equilibrium, we deduce
$$0=-R(z) +\mu_1(z-1)\alpha(z)+\nu(1-z)\frac{d\alpha(z)}{dz},
\hbox{ and }
0=+R(z) +\mu_0(z-1)\beta(z)+\nu(1-z)\frac{d\beta(z)}{dz}.$$
Adding these two relations and assuming that $z\ne 1$ gives
$$
\frac{d\alpha(z)}{dz}+\frac{d\beta(z)}{dz}=\frac{\mu_1\alpha(z)+\mu_0\beta(z)}{\nu}.
$$
This shows that the following general relations hold:
$$\E(N(\infty))=\frac{\mu_1}{\nu}P(Y(\infty)=1)+\frac{\mu_0}{\nu}P(Y(\infty)=0),$$
$${\rm Var}(N(\infty))=\frac{\mu_1}{\nu}\frac{d\alpha(z)}{dz}\vert_{z=1}+\frac{\mu_0}{\nu}\frac{d\beta(z)}{dz}\vert_{z=1}
+\E(N(\infty))(1-\E(N(\infty))),$$
where we recall that
$$\alpha(1)=P(Y(\infty)=1),\ \frac{d\alpha(z)}{dz}\vert_{z=1}=\E(N(\infty)Y(\infty))
\hbox{ and }
\frac{d\beta(z)}{dz}\vert_{z=1}=\E(N(\infty)(1-Y(\infty))).$$
These formulas make sense since, when the promoter is on (resp. off), the process evolves as a birth
and death process with birth rate $\mu_1$ (resp. $\mu_0$) and death rate $\nu n$, and has a Poisson
distribution of parameter $\mu_1/\nu$ (resp. $\mu_0/\nu$) as a stationary distribution, of mean and the variance given by $\mu_1/\nu$
(resp. $\mu_0/\nu$). The last term is related to promoter fluctuations, see the following Example.


\begin{example}\label{Constant}
Assume that  $g(n)\equiv g$
and $\kappa(n)\equiv \kappa$. Let $F(t)=\sum_{n\ge 0}nP(N(t)=n,Y(t)=1)$.
The master equation yields
$$\frac{dF(t)}{dt}=g E(t)+\mu_1 G(t)-F(t)(g+\kappa+\nu),$$
$$\frac{dE(t)}{dt}=\mu_1 G(t)+\mu_0(1-G(t))-\nu E(t),$$
where $E(t)=\E(N(t))$. 
Using the differential relation $dG(t)/dt=g(1-G(t))-\kappa G$, one gets that
$$G(\infty)=\frac{g}{g+\kappa}.$$
Notice that $G(\infty)$ is related to the stationary law of the $0/1$ Markov chain
 given by the transition rates $p(0,1)=g$ and $p(1,0)=\kappa$, which
 models the fluctuations of the state of the promoter.
Then
\begin{eqnarray*}
{\rm Var}(N(\infty))&=&\frac{d^2\alpha(z)}{d^2z}\vert_{z=1}+\frac{d^2\beta(z)}{d^2z}\vert_{z=1}+E(\infty)-E(\infty)^2\\
                    &=&\frac{\mu_1}{\nu}F(\infty)+\frac{\mu_0}{\nu}H(\infty)+E(\infty)-E(\infty)^2,
\end{eqnarray*}
where we set $H(t)=E(t)-F(t)$. It follows that
$${\rm Var}(N(\infty))=\frac{\mu_1}{\nu}G(\infty)+\frac{\mu_0}{\nu}(1-G(\infty))
+\frac{\mu_1}{\nu}F(\infty)
+\frac{\mu_0}{\nu}H(\infty)-E(\infty)^2.$$
 We finally obtain, after some algebra,
 $${\rm Var}(N(\infty))=\frac{\mu_1}{\nu}G(\infty)+\frac{\mu_0}{\nu}(1-G(\infty))
 +\frac{\tau_2}{\tau_1+\tau_2}\frac{(\mu_1-\mu_0)^2}{\nu^2}{\rm Var}(Y(\infty)),$$
 where the characteristic times $\tau_1$ and $\tau_2$ are defined by
$$\tau_1=\frac{1}{\nu}\hbox{ and }\tau_2=\frac{1}{g+\kappa}.$$
The interpretation of this formula is obtained by observing that, when
the promoter is on with probability $G(\infty)$, the process evolves as a birth and
death process with steady state distribution given by a Poisson distribution of parameter $\mu_1/\nu$. The interpretation
of the second term is similar. The third term corresponds to the variance of a Bernoulli random variable
 (on/off) multiplied by a factor accounting for characteristic times
 related to protein degradation and promoter fluctuation. 
 When $\mu_0=0$,
the coefficient of variation can be then given as
$$CV^2_N=\frac{{\rm Var}(N(\infty))}{\E(N(\infty))^2}=
\frac{1}{\E(N(\infty))}+\frac{\tau_2}{\tau_1+\tau_2}\frac{{\rm Var}(Y(\infty))}{\E(Y(\infty))^2},$$
as given in \cite{Paulsson2}. The above relations yield moreover that
$$CV^2_N=\frac{g+\kappa}{\rho g}+\frac{\nu\kappa}{g(g+\nu+\kappa)},$$
where $\rho=\mu_1/\nu$, and it follows that $CV^2_N$ is decreasing as a function
of $g$ and increasing as a function of $\kappa$.
 
\end{example}


\section{ Mean-field models\label{s.meanfield}}


 Most mathematical
works on gene networks, like \cite{Gadgil}, consider networks with linear 
transition rates in which the state space of each chemical species equals
$\N$. Results on networks involving catalytic transitions rates
are very scarce.
\cite{Lipan}, also focus on such models but allow
time dependent transition rates.
In this situation, one gets interesting linear differential equations for the first
and second moments, and for covariance functions. When some state space is finite,
 boundary effects transform the equations which become more involved.

The model for the self-regulated gene defined in
Section \ref{s.transcription.quasi} is similar to a mathematical model
for an epidemic of schistosomiasis provided by
\cite{Nasell} and \cite{Nasell2}. In their model, the authors
consider a similar Markov chain, where they replace
every external random variables in the transitions probabilities
by functions of their expected values. This means for example that
the transition rate $P((n,y),(n+1,y))=y \mu$ is replaced by
$P((n,y),(n+1,y))=\E(Y(t)) \mu$, and $P((n,y),(n,1-y))$ by
$\kappa(\E(N(t)))y+g(\E(N(t)))(1-y)$, since for this last transition, the external random variable
coresponding to this transition is $N(t)$. One gets a time-nonhomogeneous Markov
chain. One can show that the pair $(\E(N(t)),\E(Y(t)))$ converges
to a limit $(\E(N(\infty)),\E(Y(\infty)))$ (see e.g. \cite{Nasell}).
This model is then asymptotically equivalent  to the model of the self-regulated gene  given in Example \ref{Constant}
where $\mu_1=\mu_0=\E(Y(\infty))\mu$: it is easy to check that the stationary distribution
of $N(\infty)$ is Poisson of parameter $\mu_1/\nu$ when
$\nu(n)\equiv \nu n$. The mean and the variance are then equal
to $\mu_1/\nu$. The behavior of the propagation of noise
in gene networks can be counter-intuitive, as shown for
example by 
\cite{Pedraza}, where the mean gene expression
at steady state is increasing as function of some inducer, but
where the variance exhibits a peak. The same phenomenon
occurs with the therapeutic network of Section \ref{network}. 
This shows that this model can't predict
this qualitative behavior. 
We shall adopt a different point of view below by
conserving the external variable $Y(t)$ and taking only
the average of $N(t)$. 

Models in which one considers the average of $N(t)$ 
in transition rates, but not involving promoters and therefore
$Y(t)$,
have been considered more recently in the  
gene regulation setting  by
\cite{MeanField}, where the authors introduce biologically
meaningful delays in feedback interactions.
They replace occurences  in transition
rates of $\kappa(N(t))$ and $g(N(t))$  by expressions involving their expected values, that is by
$\kappa(\E(N(t-\theta)))$ and $g(\E(N(t-\theta)))$ for some delay $\theta$. 
Time delays are biologically very meaningful since,
in fact, proteins
move around at random and the delay $\theta$ might represent the average time a protein
takes to move back in the neighborhood of the promoter.
As stated in the Introduction, their models however do not involve
$Y(t)\in\{0,1\}$, and therefore promoters. For the self-regulated gene,
assuming linear degradation transition rates of the form $\nu(n)=\nu n$, the limiting
steady state is again Poisson, so that this model is not completely satisfactory for
predicting the propagation of noise in gene expression levels.

In a previous work, \cite{Fournier}, we proposed
a mean-field model, which includes promoter states and time delays, extending a model of \cite{Delayed}.
The regulatory network
was studied in living cells, and the experimental data
were in good agreement with the model's predictions.
We also provided a rationale for 
introducing mean-field interactions: The many steps and relatively
slow transitions between states of chromatin in mammalian cells
between the permissive and the non-permissive states of chromatin
may dampen the noise that stems from the stochastic binding
of a low number of activator proteins to the promoter and from noise
amplification resulting from the gene auto-activation feedback. In this
setting chromatin may act as a noise-filtering
device that allows graded response from stochastic events.
This new Markov chain $\eta(t)$ evolves in the same state space,
but has transition rates given by (we assume that $\mu_0=0$ and $\mu_1=\mu>0$)
$$q(N(t+h)=n+1,Y(t+h)=y\vert N(t)=n,Y(t)=y)=
y\ \mu \ h +o(h),$$
 $$q(N(t+h)=n,Y(t+h)=1-y\vert N(t)=n,Y(t)=y)=
 \kappa(\E(N(t-\theta))\ h+o(h)\hbox{ when }y=1,$$
 $$q(N(t+h)=n,Y(t+h)=1-y\vert N(t)=n,Y(t)=y)=
g(\E(N(t-\theta)))\ h +o(h)\hbox{ when }y=0,$$
 and
 $$q(N(t+h)=n-1,Y(t+h)=y\vert N(t)=n,Y(t)=y)=$$
 $$\nu\ n\ h+o(h).$$
 The main difference with the basic model is that
 transition rates like $p((n,y),(n,1-y))= \kappa(n)$ are
 replaced by time non-homogeneous rates
 $q_t((n,y),(n,1-y))=\kappa(\E(N(t-\theta)))$, so that the related
 Markov chain is time non-homogeneous. Let us denote by
 $Q_t$ the related transition matrix at time $t$, of
 instantaneous
 steady state distribution $\pi^{t}$, with $\pi^t Q_t=0$. In what follows,
 we shall use the family of transition matrices 
 $Q_{(b,c)}$ given by
 $$q_{(b,c)}((n,y),(n+1,y))=y\ \mu,\ q_{(b,c)}((n,y),(n-1,y))= \nu\ n,$$
 $$q_{(b,c)}((n,y),(n,1-y))= b\hbox{ when }y=1,
 \hbox{ and }
 q_{(b,c)}((n,y),(n,1-y))=c\hbox{ when }y=0,$$
 of steady state distribution $\pi^{(b,c)}$. Then
 $Q_t=Q_{(b(t),c(t))}$,
 where  $b(t)=\kappa(\E(N(t-\theta)))$ and
 $c(t)=g(\E(N(t-\theta)))$.

 When dealing with time non-homogeneous Markov chains, the main
 problem is that the law of the stochastic process
 $P(\eta(t)=(n,y))$ does not necessarily converges
 toward the limiting steady state distribution (when it exists)
 $\lim_{t\to\infty}\pi^t$, and can lead to
 oscillations, as provided for example in
 \cite{Delayed}, or in \cite{MeanField}.
 The first thing we can do is to
 check the asymptotic behavior of the functions
  $b(t)$ and $c(t)$. Suppose that these functions
  converge toward positive numbers  $b(\infty)$ and
  $c(\infty)$. Then one ask if the following holds
  \begin{equation}\label{Convergence}
  \lim_{t\to\infty}P(\eta(t)=(n,y))=\lim_{t\to\infty}\pi^t(n,y)=\pi^{(b(\infty),c(\infty))}(n,y)\ ?
  \end{equation}
  Assume that this is true: Then one gets that the steady state behavior of the
  self regulated gene is given by computing the steady state and
  the  basic statistical descriptors related to the Markov chain
  of transition kernel $Q_{(b(\infty),c(\infty))}$,
  which is much simpler. We will see that in such a situation,
  one can get exact formulas for the mean and for the variance 
  of the number of proteins (see also Example \ref{Constant}).
  
  \noindent (\ref{Convergence}) holds under fairly general assumptions.
  Theorem \ref{nonhomogeneous} of the Appendix  gives that
 $$\lim_{t\to\infty}P(\eta(t)=(n,y))=\pi^{(b(\infty),c(\infty))}(n,y),$$
 when
 the limiting process of transition kernel $Q_{(b(\infty),c(\infty))}$ is ergodic,
 \begin{equation}\label{Ergodic}
 \int_0^\infty (\sqrt{b(t)}-\sqrt{b(\infty)})^2 dt<+\infty,
 \hbox{ and }
 \int_0^\infty (\sqrt{c(t)}-\sqrt{c(\infty)})^2 dt<+\infty,
 \end{equation}
 and under an additional hypothesis which is automatically satisfied in our model
 (see the Appendix).

  The chemical master equation yields differential equations for $G(t)=P(Y(t)=1)$ and
  $E(t):=\E(N(t))$, given by
\begin{align}
 \frac{dE}{dt}=\mu G(t)-\nu E(t),\notag\\
\frac{dG}{dt}=c(t)(1-G(t))-b(t)G(t).\label{eq:SystDel2}
\end{align}

  \begin{remark} \cite{Delayed}, and \cite{MeanField}, consider
   delayed differential systems similar to the system given by
   (\ref{eq:SystDel2}) with solutions oscillating toward
  limit cycles. The steady state exists however only when
  the solutions of this system converge as $t\to\infty$. We will
  see that this is the case for a linear positive feedback.
  \end{remark}
  \medskip
  
  \centerline{\bf ALGORITHM II}
  \bigskip
  
  \indent {\bf (STEP 1):} Check the convergence of the orbits of the system given by equations (\ref{eq:SystDel2}), for a given initial condition
     $G(0)$, $E(t)$, $-\theta\le t\le 0$. When convergence holds, proceed to the next step
     
  \indent {\bf (STEP 2):} Let
$e_\infty = E(\infty)$, with

$e_\infty =\mu/\nu G(\infty).$
Solve the equation
$$g(e_\infty)(1-\frac{\nu}{\mu}e_\infty)-\kappa(e_\infty)\frac{\nu}{\mu}e_\infty=0.$$
\indent {\bf (STEP 3):} Let
$$\tau_2:=(g(e_\infty)+\kappa(e_\infty))^{-1}.$$
Compute the coefficient of variation as
 $$CV_N^2=\frac{{\rm Var}(N(\infty))}{\E(N(\infty))^2}
 =\frac{1}{e_\infty}+\frac{\tau_2}{\tau_1+\tau_2}\frac{(1-G(\infty))}{G(\infty)},$$
 where  $\tau_1=1/\nu$.
 \medskip
 
 \noindent For more insight in these formulas, see the remarks in Example \ref{Constant}.

 \subsection{Convergence for linear positive feedbacks\label{s.linear}}
 \medskip

 Recall that $c(t)=g(\E(N(t-\theta))$ and assume that $\kappa(n)\equiv\kappa$. We generalize the linear case
 by assuming that $g(x)/x$ is decreasing. 
 Notice that even when $g(x)$ is affine in $n$, this does not mean that the
 stochastic system is linear: for example, assuming fast promoters or a quasi-equilibrium, one
 gets a time nonhomogeneous birth and death process with
 birth rate $\mu c(t)/(c(t)+\kappa)$ and death rate $\nu n$.
  We prove
 below that the above
 dynamical system is such that there is 
 a globally asymptotically stable critical
 point $(E(\infty),G(\infty))$ with
 $$G(\infty)>0 \hbox{ and }c(\infty)>0,$$
 (see \cite{Gabriel}, for more general mathematical results).
 In this case, the mean and the variance of the number of proteins
 are obtained by studying the transition kernel
 $Q_{(\kappa,c(\infty))}$.

In the following, we focus on the system (\ref{eq:SystDel2}) that reads in our setting
\begin{align}
 \frac{dE}{dt}=\mu G(t)-\nu E(t),\notag\\
\frac{dG}{dt}=g(\E(N(t-\theta)))(1-G(t))-\kappa G(t),\label{eq:SystDel}
\end{align}
where $g(\cdot)$ is continuously differentiable and increasing over $\R_+$, $g(0)>0$, $\frac{g(E)}{E}$ is decreasing, the initial condition $E(t)$ is continuous and non-negative over $[-\theta,0]$ and $0\leq G(0)\leq 1$. We use the notation $\dot f$ for the derivative $df/dt$. We proceed step by step to show that condition (\ref{Ergodic}) holds.

 In Lemma \ref{lemma:existence}, we prove that the evolution equations defining the system are well defined, providing a unique solution, then we show in Lemma \ref{lemma:convergence} that the system converges to the unique biologically meaningful critical point of the system and in Lemma \ref{lemma:expconvergence} that the speed of convergence is exponential. The methods used are adapted from \cite{Gabriel}. Finally, using our hypothesis on the function $g$, it is easy to conclude that the condition (\ref{Ergodic}) holds, and the main result is stated in Theorem \ref{thrm:convnn}.\medskip\\
The theory of delayed differential equations is very different from the usual theory of differential equations, here the initial condition is no more a point in the finite dimensional space $\R_+\times [0,1]$ but a continuous nonegative function $E(t)$ over the interval $[-\theta,0]$ and a value $G(0)\in [0,1]$. To solve the system (\ref{eq:SystDel}), we have to first integrate the second equation over the interval $[0,\theta]$, then plug the solution in the first equation and integrate using the variation of constant over the interval $[0,\theta]$ and begin the whole procedure anew over the interval $[\theta,2\theta]$ with initial condition given by $E(t)$ over the interval $[0,\theta]$, and so on.
\begin{lemma}\label{lemma:existence}Existence and unicity\\
For any initial condition  $E(t)$ non-negative and continuous over $[-\theta,0]$ and $0\leq G(0)\leq 1$, there exists a unique solution of the system (\ref{eq:SystDel}) defined over $[0,+\infty)$. Furthermore, $$0<E(t)\leq\max\{E(0),\mu/\nu\}, \ t\geq 0,$$
and 
$$0<G(t)<1, \ t>0.$$ 
\end{lemma}
{\it Proof: } For any initial condition $0\leq G(0)\leq 1$ and $E(t)$ non-negative and continuous over $[-\theta,0]$, (\ref{eq:SystDel}) admits obviously a unique solution over $[0,\theta]$. If $G(0)>0$, then by continuity $G$ remains strictly positive over some open intervall to the right of $0$. If $G(0)=0$, then according to the second equation of (\ref{eq:SystDel}), $\dot G(0)>0$ and the same conclusion holds. The same reasoning shows that $G<1$ over some open intervall to the right of $0$. Clearly, if they exist, $t_0=\inf\{t\in (0,\theta],\;G(t)=0\}$ and $t_1=\inf\{t\in (0,\theta],\;G(t)=1\}$  are both strictly positive. By definition, $\dot G(t_0)\leq 0$ and by continuity, $G(t_0)=0$. The second equation of (\ref{eq:SystDel}) entails $\dot G(t_0)>0$. We have a similar contradiction for $t_1$, thus $0<G(t)<1$ over $(0,\theta]$. The variation of constant formula entails $$0<E(t)\leq\max\{E(0),\mu/\nu\}\text{ over }(0,\theta].$$ Iterating the procedure provides existence and unicity of a solution defined over $[0,+\infty)$ and the preceding inequalities are preserved.
\hfill$\square$\medskip\\
We are interested in the possible equilibria of (\ref{eq:SystDel}) in $\R_+\times [0,1]$, i.e. the solutions $(E_0,G_0)$  in $\R_+\times [0,1]$ of
$$0=\mu G_0 -\nu E_0,\quad 0=g(E_0)(1-G_0)-\kappa G_0.$$
Clearly $G_0=0$ and $G_0=1$ lead to contradictions. We thus have $0<G_0<1$ and consequently $E_0>0$. Plugging $G_0=\frac{\nu}{\mu}E_0$ in the second equation yields
$$\frac{g(E_0)}{E_0}(\frac{\mu}{\nu}-E_0)=\kappa.$$
If $g(0)>0$ and $\frac{g(E)}{E}$ is decreasing over $(0,+\infty)$, then $\frac{g(E)}{E}(\frac{\mu}{\nu}-E)$ is strictly decreasing. Since it starts at $+\infty$ and becomes ultimately negative, we conclude to the existence of a unique solution $(E_0,G_0)\in \R^2\times [0,1]$.\medskip\\
We will use the fluctuation Lemma \ref{FlucLemma} given in the Appendix to prove the convergence to the critical point $(E_0,G_0)$.
\begin{lemma}\label{lemma:convergence}Convergence\\
For any initial condition  $E(t)$ non-negative and continuous over $[-\theta,0]$ and $0\leq G(0)\leq 1$, the unique solution $(E(t),G(t))$ converges to $(E_0,G_0)$ as $t\to\infty$.
\end{lemma}
{\it Proof: } The fluctuation Lemma \ref{FlucLemma} and the monotonicity of $g$ imply that 
\begin{eqnarray}
&0\geq \mu\underline G -\nu \underline E,
&\quad 0\geq g(\underline E)(1-\underline G)-\kappa\underline G\notag\\
&0\leq \mu\overline G -\nu \overline E,
&\quad 0\leq g(\overline E)(1-\overline G)-\kappa\overline G.\label{eq:fluc}
\end{eqnarray}
We prove the last inequality to exemplify the method. We choose $t_n\uparrow+\infty$ so that $G(t_n)\rightarrow \overline G$ and $ \dot G(t_n)\rightarrow0$ as $n\rightarrow+\infty$. Since the sequence $\{E(t_n)\}_{n\geq 1}$ is bounded, there exists a subsequence $(t_{n_k})_{k\geq 1}$ so that $E(t_{n_k})$ converges as $k\to \infty$ to a certain value that we call $E_\infty$. Evaluating the equation for $G$ over the subsequence $(t_{n_k})_{k\geq 1}$ and letting $k\rightarrow+\infty$, we get
$$0=g(E_{\infty})(1-\overline G)-\kappa \overline G\leq g(\overline E)(1-\overline G)-\kappa \overline G$$
since $g$ is increasing and $\overline G\leq 1$. The proof of the other inequalities in (\ref{eq:fluc}) is similar.\\
We already know that $0<\underline G$, $\underline E>0$ and $\overline G<1$, and (\ref{eq:fluc}) entails  $\underline G\leq\frac{\nu}{\mu}\underline E$ and $\overline G\geq\frac{\nu}{\mu}\overline E$, and in particular $\overline E\leq \frac{\mu}{\nu}\overline G<\frac{\mu}{\nu}$. Consequently
$$\frac{\kappa\nu}{\mu}\overline E\leq\kappa\overline G\leq g(\overline E)(1-\overline G)\leq g(\overline E)(1-\frac{\nu}{\mu}\overline E),$$
 hence 
$$\kappa\leq\frac{g(\overline E)}{\overline E}(\frac{\mu}{\nu}-\overline E).$$
Repeating the same argument for $\underline E$, one gets
$$\kappa\geq\frac{g(\underline E)}{\underline E}(\frac{\mu}{\nu}-\underline E).$$
By assumption, $\frac{g(E)}{E}(\frac{\mu}{\nu}-E)$ is decreasing, so that the two last equations then give that $\overline E\leq E_0$ and $\underline E\geq E_0$. Clearly we have $\underline E=\overline E= E_0$, so that $E(t)$ converges as $t\rightarrow+\infty$. According to Lemma $3.1$ in \cite{Coppel},  $E(t)$ and its first two derivatives being bounded on $[\theta,+\infty)$,  we have $\lim\limits_{t\rightarrow +\infty}\dot E(t)=0$ and the relation $\dot E=\mu G-\nu E$ entail the convergence of $G(t)$ as $t\rightarrow+\infty$.
\hfill$\square$\medskip\\
From this Lemma, we deduce that the propensity function $$a_3(t,y)=g(\E(N(t-\theta)))\ y$$ converges to $g_\infty\ y=g(E_0)\ y$ as $t\to\infty$. To show that the convergence speed is exponential, we use Theorem \ref{expopoly} cited in the Appendix.
\begin{lemma}\label{lemma:expconvergence}Exponential convergence\\
The convergence of $(E(t),G(t))$ to $(E_0,G_0)$ is exponential.
\end{lemma}
{\it Proof: } Near a critical point, the asymptotic behaviour of the system is determined by the asymptotic behaviour of the linearized system
\begin{equation*}\label{linearized}\left[\begin{array}{c}\dot E(t)\\\dot G(t)\end{array}\right]=A\cdot \left[\begin{array}{c} E(t)\\ G(t)\end{array}\right]+B\cdot \left[\begin{array}{c} E(t-\theta)\\ G(t-\theta)\end{array}\right],\end{equation*}
where $A$ and $B$ are the matrices
$$A=\left[\begin{array}{cc}-\nu & \mu \\0 & -(g(E_0)+\kappa)\end{array}\right],\qquad B=\left[\begin{array}{cc}0 & 0 \\g'(E_0) (1-G_0) & 0\end{array}\right].$$
We show that all roots $\lambda$ of the 
caracteristic equation 
$\det(A+e^{-\lambda \theta}B-\lambda I)=0$ have negative real parts. The caracteristic equation is here
$$\lambda^2+\lambda (\nu+g(E_0)+\kappa)+\nu (g(E_0)+\kappa)-\mu g'(E_0)(1-G_0)e^{-\lambda \theta}=0,$$
and all roots $\lambda$ of this equation have negative real part if and only if all roots $z$ of
$$H(z):=(z^2+pz+q)e^z+r=0$$
have negative real parts, 
with $$p:=(\nu+g(E_0)+\kappa)\theta, \quad q:=\nu (g(E_0)+\kappa)\theta^2, \quad r:=-\mu g'(E_0)(1-G_0)\theta^2,$$
and the change of variable $z:=\lambda\theta$. According to Theorem \ref{expopoly} given in the Appendix, since $r<0$ and $$p^2=2q+\nu^2\theta^2+(g(E_0)+\kappa)^2\theta^2\geq 2q,$$ we have to check that $-q<r<0$ and $r\sin(a_2)/(pa_2)\leq 1$, where $a_2$ is the unique root of the equation $\cot(a)=(a^2-p)/q$ which lies in the interval $(2\pi,3\pi)$. The second inequality is clear since $r/p<0$ and $\sin(x)/x\geq 0$ on $(2\pi,3\pi)$. For the inequality $-r<q$, notice that $g'(x)\leq g(x)/x$ since $g(x)/x$ is decreasing,  and using the equilibrium equation $g(E_0)(1-G_0)=\kappa G_0=\frac{\nu\kappa}{\mu}E_0$, we have
$$-r=\mu  g'(E_0)(1-G_0)\theta^2\leq \mu\frac{g(E_0)}{E_0}(1-G_0)\theta^2
=\nu\kappa \theta^2<\nu (\kappa+g(E_0))\theta^2=q.$$
Hence all roots of the characteristic equation have negative real parts and the system is asymptotically stable.  Since our system is autonomous, asymptotic stability implies uniform asymptotic stability. According to theorem 4.6 in \cite{Halanay}, the convergence is exponential.
\hfill$\square$\medskip\\
Using the exponential convergence of $\E(N(t-\theta))$ and the hypothesis on $g$, it is now easy to show that condition (\ref{Ergodic}) is satisfied.
\begin{theorem}\label{thrm:convnn} Assume that $g(n)=g_0+g_1 n$ with $g_0>0$ and that
$\kappa(n)\equiv \kappa$. The limiting
distribution of the time-nonhomogeneous process $(N(t),Y(t))$ is such that
$$\lim_{t\to\infty}P(N(t)=n,Y(t)=y)=\pi_n^{(\kappa,g(E_0))}(y),$$
where $\pi^{(g(e_\infty),\kappa)}$ is the steady state distribution given
by Theorem \ref{Steady} for a self regulated gene with the simpler transitions 
$$P((n,y),(n+1,y))=y\ \mu,\ \ P((n,y),(n-1,y))=\nu n,$$
 $$P((n,y),(n,1-y))=\kappa \hbox{ when }y=1,\ \ 
 P((n,y),(n,1-y))=g(E_0)\hbox{ when }y=0.$$
The steady state coefficient of variation
of $N(\infty)$ is given by
$$CV_N^2=\frac{1}{E_0}+\frac{\tau_1}{\tau_1+\tau_2}\frac{\kappa}{g(E_0)},$$
where 
$$\tau_2=\frac{1}{g(E_0)+\kappa},\ \ \tau_1=\frac{1}{\nu}.$$
\end{theorem}
{\it Proof: } According to Theorem \ref{nonhomogeneous} in the Appendix, we only have to show that condition (\ref{Ergodic}) holds. Using the positiveness and boundedness of $g(E(t-\theta))$, $0<g(0)\leq g(E(t-\theta))\leq g(\mu/\nu)$, and the expansion
$$\sqrt{g(E(t-\theta))}-\sqrt{g(E_0)}=\frac{g(E(t-\theta))-g(E_0)}{\sqrt{g(E(t-\theta))}+\sqrt{g(E_0)}},$$ we have
$$\frac{\mid g(E(t-\theta))-g(E_0)\mid}{2\sqrt{g(\mu/\nu)}}\leq \mid\sqrt{g(E(t-\theta))}-\sqrt{g(E_0)} \mid\leq \frac{\mid g(E(t-\theta))-g(E_0)\mid}{2\sqrt{g(0)}}$$
and condition (\ref{Ergodic}) is in our case equivalent to $$ \int_0^\infty \big(g(E(t-\theta))-g(E_0)\;\big)^2 dt<\infty.$$ Let $\varepsilon$ be positive, $\varepsilon<E_0$ and $T_{\varepsilon}$ be such that $\mid E(t-\theta)-E_0\mid<\varepsilon$ for all $t\geq T_{\varepsilon}$. Using the mean value theorem, for all $t\geq T_{\varepsilon}$, there exists a $\xi_t$ in the interval delimited by $E(t-\theta)$ and $E_0$ such that
$$\mid g(E(t-\theta))-g(E_0)\mid =g'(\xi_t)\ \mid E(t-\theta)-E_0\mid.$$
Since for all $t\geq T_{\varepsilon}$, $E_0-\varepsilon<\min(E(t-\theta),E_0)$, and furthermore $0\leq g'(x)\leq g(x)/x$ and $g(x)/x$ is decreasing, 
\begin{align*}
\mid g(E(t-\theta))-g(E_0)\mid &=g'(\xi_t)\ \mid E(t-\theta)-E_0\mid \leq \frac{g(\xi_t)}{\xi_t}\mid E(t-\theta)-E_0\mid \\
&\leq \frac{g(E_0-\varepsilon)}{E_0-\varepsilon}\ \mid E(t-\theta)-E_0\mid
=:L_{\varepsilon}\mid E(t-\theta)-E_0\mid,
\end{align*}
and finally with the exponential convergence of $E(t-\theta)$ to $E_0$, condition (\ref{Ergodic}) holds
\begin{align*}
\int_0^\infty \big(g(E(t-\theta))-g(E_0)\big)^2 dt \leq &\displaystyle \int_0^{T_{\varepsilon}} \big(g(E(t-\theta))-g(E_0)\big)^2 dt\\ &+L_{\varepsilon} \displaystyle \int_{T_{\varepsilon}}^\infty \hspace{-0.1 cm}\big(E(t-\theta)-E_0\big)^2 dt<\infty.
\end{align*}
\hfill$\square$\medskip\\

\begin{remark}When the positive feedback rate $g(\E(N(t-\theta)))$ is such that $\frac{g(x)}{x}$ is increasing, for example when $g$ is a polynomial of degree $\geq 2$, there can possibly exist several biologically meaningful equilibrium points and it can not be excluded that for some initial conditions the solutions of equation (\ref{eq:SystDel}) oscillate endlessly.\medskip\\
When $g$ is constant but the negative feedback $\kappa(\E(N(t-\theta)))$ is an increasing function of $\E(N(t-\theta))$, the biologically meaningful equilibrium point is unique but similar application of the fluctuation lemma as in the proof of Lemma \ref{lemma:convergence} yields the trivial observation that $\underline{E}\leq E_0\leq \overline{E}$, and oscillating solutions can not be excluded in this case either.
 \end{remark}

\section{ Two-time-scale stochastic simulations\label{s.two}}

The self-regulated gene and the network presented in the
Introduction
 involve slow
and fast species, like therapeutic
proteins and activator dimers. 
We recall existing known probabilistic
results concerning quasi-equilibrium.
Let $\ep >0$ be a small parameter, which
will be useful for describing fast species.
In what follows, $\eta^\eps(t)$ is a random
vector describing the number of molecules of each
species present in the cell at time $t$.
For example, considering the self-regulated gene,
$N^\eps (t)$ gives the number of protein molecules at time $t$, and
$Y^\eps(t)=0,\ 1$ gives the state of the promoter. The pair
$\eta^\eps_s(t)=(N^\eps(t),Y^\eps(t))$ stands for the slow
process. $\eta^\eps_f(t)$ models the fast process, and the
global process is $\eta^\eps(t)=(\eta^\eps_s(t),\eta^\eps_f(t))$.

A generic example of fast reaction is dimerization, as given by the
chemical reaction
$${\cal M}+{\cal M}\underset{\beta_-^{\varepsilon} }{\overset{\beta_+^{\varepsilon}}{\longleftrightarrow}}{\cal D},$$
where ${\cal M}$ represents protein monomers and ${\cal D}$ protein dimers. 
Protein dimers form a fast species, while protein (involved in dimers or monomers)
is a slow species.
The rates of these reactions
are fast when for example the rate constants $\beta^\eps_-$ and $\beta^\eps_+$ are such that
$\beta_-^\eps=c_-/\eps$ and $\beta_+^\eps=c_+/\eps$, 
for positive constants $c_-$ and $c_+$, when $\eps\approx 0$. In this setting, the number of protein
monomers is then given by $N^\eps(t)-2D^\eps(t)$, where $D^\eps(t)$
gives the number of protein dimers present in the cell at time $t$.
Here
$$\eta^\eps_s(t)=(N^\eps(t),Y^\eps(t))\hbox{ and }\eta^\eps_f(t)=D^\eps(t).$$
When $\eps\approx 0$, a quasi-equilibrium is attained, meaning that
for given $\eta^\eps_s(t)=k$, one can assume a local steady state
for the number of dimers. 
More generally,
we assume that the slow process evolves in some finite space
$E_s=\{1,\cdots,L\}$, and that, given $k\in E_s$,
$ \eta_f^\eps(t)\in\N$ is described by a Markov transition kernel
$A^k(t)/\eps$ (see below).
We follow essentially \cite{Zhang}.
We assume that the generator $Q(t)=Q^{\ep}(t)$ of the
Gillespie algorithm $\eta^\eps(t)$ can be decomposed as
$$
Q^{\ep}(t)=\frac{1}{\ep} A(t) +B(t),
$$
where $A(t)$ and $B(t)$ are matrix valued functions. Following \cite{Zhang}, assume
that
$A(t)$ has the block diagonal form
$$A(t)=\begin{pmatrix}A^1(t)&0&0&\cdots&0\cr
                 0&A^2(t)&0&\cdots&0\cr
                 \cdots&\cdots&\cdots&\cdots\cr
                 0&0&0&\cdots&A^L(t)\end{pmatrix},$$
where each block $A^k(t)/\ep$ is a transition matrix representing the transitions rates
of the fast variables given $ \eta^\eps_s=k$.  The generator $B(t)$ gives the slow
transition rates and in particular  transitions of the form
$((k,u),(j,v))$, where $u,v\in \N$ and $k,j\in E_s$.  Following \cite{Zhang},
we partition the state space $E$ as
$$E=E_1\cup E_2\cup\cdots\cup E_L,$$
where each $E_k$ contains $m_k$ elements, with
$$E_k=\{e_{k1},e_{k2},\cdots,e_{km_k}\},\ k\in E_s.$$
Each $E_k$ corresponds to some subset
of  $\{k\}\ {\rm x}\ \N$, with $k\in E_s$.

\bigskip

\noindent {\bf Hypothesis}:
We suppose that the process is time homogeneous, that is that both $A(t)$ and $B(t)$ do not
depend on $t$, and that each generator $A^k$ is irreducible with a unique invariant probability
measure $\sigma^k=(\sigma^k(e_{k1}),\cdots,\sigma^k(e_{km_k}))$, such that
$\sigma^k A^k=0$.
\bigskip

Following \cite{Zhang}, each $E_k$ can be aggregated, and represented by a single state $k$,
corresponding to a particular slow state; The Markov process $ \eta^\varepsilon(t)$
of transition kernel $Q^\varepsilon$ is then approximated by an {\it aggregated process}
$\bar \eta^\varepsilon(t)$ defined by
$$\bar \eta^\varepsilon(t)=k\hbox{ if } \eta^\varepsilon(t)\in E_k,\ \ k=1,\cdots,L.$$
This process converges in distribution as $\varepsilon\to 0$
toward a Markov process $\eta(t)$ generated
by the kernel
 $ Q=(\gamma_{kj})_{k,j\in E_s}$, with
\begin{equation*}\label{LimitingKernel}
\gamma_{kj}=\sum_{u=1}^{m_k}\sum_{v=1}^{m_j} \sigma^k(e_{ku})B(e_{ku},e_{jv}),\ k\ne j.
\end{equation*}

\subsection{Transcription with fast dimerization\label{s.transcription.dimerization}}
The model is similar to that given in Section \ref{s.transcription.quasi}, with dimerization
as a fast component, see e.g. \cite{Burrage}, \cite{Cao}, or \cite{Goutsias}. It is described by the following
set of chemical reactions:
$$
{\cal M}{\overset{\nu }{\longrightarrow}}\emptyset,\ \ 
\emptyset{\overset{\mu_l }{\longrightarrow}}{\cal M},\ l\in\{0,1\},\ 
{\cal O}_0+{\cal D} \underset{\kappa(d) }{\overset{ g(d)}{\longleftrightarrow}}{\cal O}_1,\ 
{\cal M}+{\cal M}\underset{\beta_-^{\varepsilon} }{\overset{\beta_+^{\varepsilon}}{\longleftrightarrow}}{\cal D},
$$
where ${\cal D}$ represent dimers, $ g$ is function of the number of dimers, and 
the rates $\beta_-^{\varepsilon}$ and $\beta_+^{\varepsilon}$ involve a small number $\varepsilon >0$
modeling the speed of dimerization, see below. The number $N^\ep(t)$ of proteins ${\cal P}$ 
present at time $t$ is related to the number of dimers as $0\le 2D^\ep(t)\le N^\ep (t)$, and
the number of free monomers is such that ${\cal M}+2 {\cal D}={\cal P}$.

The running process is  a Markov process
$ \eta^\varepsilon (t)=(N^\varepsilon (t),Y^\varepsilon (t),D^\varepsilon (t))$, $0\le 2 D^\varepsilon (t)\le N^\varepsilon (t)$, $t\ge 0$.
The dimerization process $D^\varepsilon (t)$ is given by the transition rates 
$$P(\eta_s^\ep(t+h)=\eta_s^\ep(t),D^\varepsilon(t+h)=D^\varepsilon(t)+1)
=\beta_+^{\varepsilon}(N^\varepsilon(t)-2D^\varepsilon(t))(N^\varepsilon(t)-2D^\varepsilon(t)-1) h+o(h),$$
$$P(\eta_s^\ep(t+h)=\eta_s^\ep(t),D^\varepsilon(t+h)=D^\varepsilon(t)-1)=\beta_-^{\varepsilon} D^\varepsilon(t) h+o(h),$$
where the slow process is $\eta^\varepsilon_s(t)=(N^\varepsilon(t),Y^\varepsilon(t))$.

The transition rates of $D^\varepsilon (t)$ depend on $N^\varepsilon (t)$ but are independent of
the state of the promoter. We can fit the setup of this Section 
by setting
\begin{equation*}\label{rates}
\beta_-^{\varepsilon}=\frac{c_-}{\varepsilon}\hbox{ and }\beta_+^{\varepsilon}=\frac{c_+}{\varepsilon},
\end{equation*}
for positive constants $c_-$ and $c_+$. Then, the result of Section \ref{s.two}
yield that the slow process at quasi-equilibrium ($\eps\approx 0$) is well described
in the above discussion: For a given slow state
  $k=(n,y)$, one gets
$$g(n)=\sum_{0\le d\le [n/2]}\sigma^{(n,y)}(d) g(d),\ \ \sigma^{(n,y)}=\mu^n,$$
 where the quasi-equilibrium stationary measure $\sigma^{(n,y)}$  corresponds to the stationary measure $\mu^n$ of the
 dimerization process (see below).
A typical example is given by
$ g(D^\varepsilon (t))=\lambda D^\varepsilon (t)+g(0)$, that is depends linearly on the number
of dimers at time $t$. Then,  at quasi-equilibrium one gets
 $$
 g(n)=\sum_{0\le d\le [n/2]}\mu^n (d)(\lambda d+g(0))=\lambda \E_n+g(0),\ \hbox{ where we set } \E_n:=\sum_{0\le d\le [n/2]}d \mu^n(d),
$$
where $[\cdots]$ denots the integer part.
The algorithms developped in Section \ref{s.transcription.quasi} can be applied efficiently if one
can compute the rates $ g(n)$. The next Section develops efficient
algorithms for computing $ g(n)$ when $ g(d)$ is linear or quadratic in the number $d$ of dimers.

\subsection{ Dimerization\label{s.dimerization}}

Dimerization appears in most biochemical processes, and is usually
considered as a fast reaction. The aim of this Section is to give
mathematical statements relevant for computational purposes
(see also \cite{Darvey}, \cite{Cao}, or \cite{Kepler}). Given a fixed number  of proteins $n$,
the dimerization process is given by the reaction
$${\cal M}+{\cal M}\underset{c_-}{\overset{c_+ }{\longleftrightarrow}}{\cal D},$$
where we recall that ${\cal M}$ and ${\cal D}$ represent protein monomers and dimers.
 The
infinitesimal transitions probabilities are such that
$$P(D(t+h)=i+1|D(t)=i)=c_+(n-2i)(n-2i-1)\,h+o(h),$$
$$ P(D(t+h)=i-1|D(t)=i)=c_-\,i\,h+o(h).$$
The stationary distribution $\mu^n$ of the process is given explicitely by
\begin{equation*}\mu^n(i)= \Big(\frac{c_+}{ c_-}\Big)^i\frac{1}{(n-2i)!\,i!}\cdot\frac{1}{Z_n},\ 0\le i\le n_2,\ n_2:=[n/2],\end{equation*}
where
\begin{equation*}\label{Z}Z_n =\sum\limits_{i=0}^{n_2}\Big(\frac{2c_+}{ c_-}\Big)^i\frac{1}{(n-2i)!\,i!\cdot 2^i}.\end{equation*}

It can be shown that the generating function at equilibrium $M(s)=\sum\limits_{i=0}^{n_2}\mu^n(i)s^i$ can be written using confluent hypergeometric functions
\begin{equation*}M(s)=\begin{cases}s^{n_2}\frac{{}_1F_1\big(-n_2,\frac{3}{2},-\frac{c_-}{4c_+s}\big)}{{}_1F_1\big(-n_2,\frac{3}{2},-\frac{c_-}{4c_+}\big)}&\text{ if }n \text{ is odd, }\\
&\\
s^{n_2}\frac{{}_1F_1\big(-n_2,\frac{1}{2},-\frac{c_-}{4c_+s}\big)}{{}_1F_1\big(-n_2,\frac{1}{2},-\frac{c_-}{4c_+}\big)}&\text{ if }n \text{ is even. }\end{cases}
\end{equation*}
This gives a theoretical way of computing the invariant measure as 
$$\mu^n(i)=\frac{M^{(i)}(0)}{i!},\qquad 0\leq i \leq n_2,$$
and the mean number of dimers in the stationary regime is given by
\begin{equation*}\E_n=M'(1).\end{equation*}
Numerical computation based on this last formula is tedious and in the case described in Section \ref{s.networkquasi} we have to compute moments repeatedly for each $n$ between $0$ and $\Lambda$. The recursive method described in the next section provides an alternative adapted to this situation.

\subsection{An approach of the invariant measure adapted to numerical computation}

We provide a different approach, which will allow efficient computations of  the mean and second moment. 
If the feedback function in the slow process is linear or quadratic in the number of dimers, 
the infinitesimal transition rates of the slow process at quasi-equilibrium will only 
depend on the first two moments.
Set $y =c_+/c_-$, so that $\displaystyle Z_n=\sum\limits_{i=0}^{n_2}\frac{y^i}{(n-2i)!\,i!}$. Using the following polynomial identities :
$$\sum_{i=1}^{n_2}\frac{iy^i}{(n-2i)!i!}=yZ_{n-2}\quad\text{ and }\quad
\sum_{i=1}^{n_2}\frac{i^2y^i}{(n-2i)!i!}=y^2 Z_{n-4}+yZ_{n-2},
$$
the mean $\E_n:=\sum_{0\le i\le n_2}i\mu^n(i)$ and second moment $\E_n^2:=\sum_{0\le i\le n_2}i^2\mu^n(i)$ are given by
\begin{equation}\label{EZ}\E_n=\frac{1}{Z_n}\sum_{i=1}^{n_2}\frac{iy^i}{(n-2i)!i!}=y\frac{Z_{n-2}}{Z_n}
\end{equation}
\begin{equation*}
\E_n^2=\frac{1}{Z_n}\sum_{i=1}^{n_2}\frac{i^2y^i}{(N-2i)!i!}=y^2 \frac{Z_{n-4}}{Z_n}+y\frac{Z_{n-2}}{Z_n}\notag
=\E_n\big(1+\E_{n-2}\big). \notag 
\end{equation*}
In what follows, we give another description of $Z_n$ based on the involutions of the permutation group $S_n$. This approach will allow to compute the ratios $Z_{n-2}/Z_n$ recursively.


\subsubsection{A description of $Z_n$ based on the involutions of $S_n$}


Let $I_n$ denote the involution subgroup of the permutation group $S_n$, i.~e. the set of permutation of $n$ points $\sigma \in S_n$ so that $\sigma^2$ is the identity. For $\sigma \in I_n$,  $\text{fix}(\sigma)$ denotes the number of fixed points of $\sigma$. For any  number $n-2i$ between $0$ and $ n$, the number of involutions with $n-2i$ fixed points is given by
$$\underset{\text{fix}(\sigma)=n-2i}{\sum_{\sigma\in I_n}}1 =\binom{n}{2}\cdot\binom{n-2}{2}\cdots\binom{n-2(i-1)}{2}\cdot\frac{1}{i!}= \frac{n!}{(n-2i)!\,i!\cdot 2^i},$$
so that, setting $\displaystyle q:=\sqrt{c_-/2c_+}=(2y)^{-1/2}$, $Z_n$ can be written as
$$Z_n =\sum\limits_{i=0}^{n_2}\Big(\frac{2c_+}{ c_-}\Big)^i\frac{1}{(n-2i)!\,i!\cdot 2^i}
=\frac{1}{n!}\sum_{\sigma \in I_n}\Big(\frac{2c_+}{ c_-}\Big)^{\frac{n-\text{fix}(\sigma)}{2}}
=\frac{q^{-n}}{n!}\sum_{\sigma \in I_n}q^{\text{fix}(\sigma)}$$
Let $Q_n$ denote the polynomial
$\displaystyle Q_n(q):=\sum_{\sigma \in I_n}q^{\text{fix}(\sigma)},$
so that the partition function and the mean (\ref{EZ}) can be written as
\begin{equation}Z_n=\frac{q^{-n}}{n!}Q_n(q),\quad\text{ and }\quad
\E_n=\frac{n(n-1)}{2}\frac{Q_{n-2}(q)}{Q_n(q)}.\label{mean1}\end{equation}
According to \cite{Rand}, one can identify $Q_n$ as the Taylor coefficient of a Stieltjes type continued fraction. Here we proceed in a recursive way, using the following two propositions.
\begin{proposition} \label{prop1}$Q_n(q)$ satisfies the relation $Q_{n+1}(q)=q\,Q_n(q)+Q_n'(q).$
\end{proposition}
{\bf Proof} \; Each involution $\sigma \in I_n$ induces $1+\text{fix}(\sigma)$ involutions in $I_{n+1}$, namely the one that  fixes the point $n+1$ and the $\text{fix}(\sigma)$ ones that interchange a fixed point of $\sigma$ with $n+1$. 
Partitioning $I_{n+1}$ as the set of involutions that fixe  $n+1$, and those that do not, we see that the first set contains exactly the involutions of $I_n$ except that they have one more fixed point, namely $n+1$. For each $\sigma\in I_n$ that fixes at least one point, the second set contains $\text{fix}(\sigma)$ involutions with one fixed point less, namely the one that is interchanged with $n+1$. More precisely, the partition of $I_{n+1}$ is given by 
\begin{align*}
I_{n+1}&=\{\sigma \in I_{n+1} ; \sigma \text{ fixes } n+1\}\cup \{\sigma \in I_{n+1} ; \sigma \text{ does not fix } n+1\}\\
&=I_n\cup\bigcup_{\sigma\in I_n}\;\bigcup_{\sigma\text{ fixes }k}\{\sigma\circ (k,n+1)\}\end{align*}
where $(k,n+1)$ is the permutation of $k$ and $n+1$. Therefore, we have the recurrence relation
\begin{align*}
Q_{n+1}(q)&=\sum_{\sigma\in I_{n+1}}q^{\text{fix}(\sigma)}=\sum_{\sigma\in I_{n}}q^{\text{fix}(\sigma)+1}+\underset{\text{fix}(\sigma)\geq 1}{\sum_{\sigma\in I_{n},}}\sum_{1\leq k\leq \text{fix}(\sigma)}q^{\text{fix}(\sigma)-1}\\
&=q \sum_{\sigma\in I_{n}}q^{\text{fix}(\sigma)}+\underset{\text{fix}(\sigma)\geq 1}{\sum_{\sigma\in I_{n},}}\text{fix}(\sigma)q^{\text{fix}(\sigma)-1}=q\,Q_n(q)+Q_n'(q).
\end{align*}
$\hfill\square$

\begin{proposition} \label{prop2} The derivative of $Q_n$ is given by $Q_n'(q)=n\cdot Q_{n-1}(q)$, and hence \begin{equation}\label{rec}Q_{n+1}(q)=q\,Q_n(q)+n\cdot Q_{n-1}(q).\end{equation}
\end{proposition}
{\bf Proof} \; One can easily compute $Q_1(q)=q$ and $Q_2(q)=q^2+1$. If  $Q_n'(q)=nQ_{n-1}(q)$ for some $n$, using Proposition \ref{prop1} for the first and last equality and by the induction hypothesis for the second one, we have  $$Q_{n+1}'(q)=Q_n(q)+qQ_n'(q)+Q_n''(q)=Q_n(q)+qnQ_{n-1}(q)+nQ_{n-1}'(q)=(n+1)Q_n(q).$$
$\hfill\square$
\begin{remark} Let  $h_q(t):=\sum_{n=0}^{\infty}Q_n(q)\frac{t^n}{n!}$.
Multiplying both sides of (\ref{rec}) by $\displaystyle \frac{t^n}{n!}$ and summing over all possible $n$ leads to the equation
$h'_q(t)=(t+q)h_q(t)$,
with initial condition $h_q(0)=Q_0(q)=1$, which has the unique solution
$$\label{hq}h_q(t)=e^{\frac{t^2}{2}+qt}.$$
This function is  the moment generating function of a normal random variable  of mean $q$ and variance $1$, so that $Q_n(q)$ is the $n$-th moment of a random variable $X \sim{\mathcal N}(q,1)$.
\end{remark}

\bigskip

\centerline{\bf ALGORITHM III}
\bigskip

Due to the fast increase of its coefficients, $Q_n$ cannot be efficiently computed for large $n$. However, the computation of the mean $\E_n$ only involves the ratio $Q_{n-2}/Q_n$.\\ 
Let  $\displaystyle c_n(q):=\frac{Q_{n-1}(q)}{Q_n(q)}$.
From
$\displaystyle \frac{Q_{n-2}(q)}{Q_n(q)}=c_{n-1}(q)\cdot c_n(q)=\frac{1}{n-1}\big(1-q\,c_n(q)\big),$
 one obtains the continued fraction
\begin{equation}\label{reccN}
 \frac{1}{c_{n+1}(q)}=q+n c_n(q).
\end{equation}
From $Q_0(q)=1$ and $Q_1(q)=q$, the first term $c_1$ is given by $c_1(q)=1/q$.\\
The mean and the second moment (\ref{mean1}) or (\ref{EZ}) can then be computed recursively as
\begin{align*}\E_n&=\frac{n}{2}\big(1-q\,c_n(q),\big)\\
\E_n^2&=\frac{n}{2}\big(1-q\,c_n(q)\big)\Big(1+\frac{n-2}{2}\big(1-q\,c_{n-2}(q)\big).\Big)
\end{align*}
 
\begin{theorem}\label{prop:ctozero}
$c_n(x)\to 0$ as $n\to \infty$.
\end{theorem}
{\it Proof:} Suppose that the $\limsup$ of the sequence of non-negative numbers $\{c_n(x)\}_{n \geq 1}$ is strictly positive, $$\limsup_{n\to\infty}c_n(x)=a>0.$$ Using relation $(\ref{reccN})$, one gets
$$ a=\limsup_{n\to\infty}c_n(x)=\frac{1}{x+\liminf\limits_{n\to\infty}nc_n(x)}=:\frac{1}{x+b},$$
where $\displaystyle b:=\liminf_{n\to\infty}nc_n(x)$ has to be finite. Isolating $c_n(x)$ in $(\ref{reccN})$ yields $$c_n(x)=\frac{1}{nc_{n+1}(x)}-\frac{x}{n}$$ and we get 
$$a=\limsup_{n\to\infty}c_n(x)=\limsup_{n\to\infty}\left(\frac{1}{nc_{n+1}(x)}-\frac{x}{n}\right)=\frac{1}{\liminf\limits_{n\to\infty}nc_{n+1}(x)\cdot\frac{n+1}{n+1}}=\frac{1}{b}.$$
Since $x>0$ and $b<\infty$, this leads to the contradiction $\displaystyle \frac{1}{x+b}=\frac{1}{b}$.
\hfill$\square$\medskip\\

The above Theorem leave to the somehow counterintuitive conclusion that the fraction of dimers is about $\frac{1}{2}$ for $n$ large, more precisely $$\lim_{n\to \infty}\frac{\E_n}{n}=\frac{1}{2},$$
for every set of positive parameters $c_+,\ c_-$.

In our concrete Example of Section \ref{s.transcription.dimerization}, we are mainly interested in computing higher moments for $n$ proteins. We will show below that the computation of higher moments only requires the knowledge of the first moments for a lower number of proteins. More precisely, let  $P_{j+1}(i)$ denote the polynomial
\begin{equation*}\label{pj}P_{j+1}(i):=i\cdot(i-1)\cdots(i-2)\cdots(i-j)=:i^{j+1}-\sum_{l=1}^j a_{l,j} i^l.\end{equation*}
With the convention that $\E_i=0$ for $i<0$, the higher moments can be computed as combinations of the means for lower total number of proteins.
\begin{lemma}\label{l:hmom}
$\E_n(P_{j+1}(D))=\E_{n-2j}\cdot\E_{n-2(j-1)}\cdots\E_{n-2}\cdot\E_n.$
\end{lemma}
{\it Proof:} We show that both terms are equal to $\displaystyle y^{j+1}\ \frac{Z_{n-2(j+1)}}{Z_n}.$
\begin{align*}Z_n\cdot\E_n(P_{j+1}(D))=&\sum_{i=1}^{n_2}\frac{P_{j+1}y^i}{(n-2i)!\ i!}=\sum_{i=1}^{n_2}\frac{i\cdot(i-1)\cdots(i-2)\cdots(i-j)\ y^i}{(n-2i)!\ i!}\\
=&\sum_{i=1}^{n_2}\frac{y^i}{(n-2i)!(i-j-1)!}\cdot 1_{\{i>j\}}\\
=&\sum_{i=j+1}^{n_2}\frac{y^i}{(n-2(i-j-1)-2(j+1))!\ (i-j-1)!}\\=&\sum_{i=0}^{n_2-(j+1)}\frac{y^{i+j+1}}{(n-2(j+1)-2i)!\ i!}\\
=&y^{j+1}Z_{n-2(j+1)},\end{align*}
and with (\ref{mean1}), we have
\begin{align*}
\E_{n-2j}\cdot\E_{n-2(j-1)}\cdots\E_{n-2}\cdot\E_n=&y\ \frac{Z_{n-2(j+1)}}{Z_{n-2j}}\cdot y\ \frac{Z_{n-2j}}{Z_{n-2(j-1)}}\cdots y\ \frac{Z_{n-2}}{Z_n}\\
=&y^{j+1}\ \frac{Z_{n-2(j+1)}}{Z_n}.\qquad\qquad \qquad\qquad \quad\quad \square
\end{align*}

From the preceding Lemma, we can give a formula for arbitrary moments:
\begin{theorem}\label{mom}The $j+1$-th moment of $D$ is given by
\begin{equation*}\E_n^{j+1}=\E_{n-2j}\cdot\E_{n-2(j-1)}\cdots\E_{n-2}\cdot\E_n+\sum_{l=1}^j a_{l,j} \E_n^l.\end{equation*}\end{theorem}
{\it Proof:} From the definition of the coefficients $a_{l,j}$, 
$$\E_n(P_{j+1}(D))=\E_n^{j+1}-\sum_{l=1}^j a_{l,j} \E_n^l,$$
hence with Lemma \ref{l:hmom} the statement holds.

\section{Modeling the regulatory gene network\label{network}\label{Network}}
We first recall the basic mathematical steps which lead to the mathematical model studied in
\cite{Fournier}.
We shall see that the time evolution of the gene products involved
in the network described in the Introduction can be modeled by the following set
of chemical reactions:

$$
{\cal A}{\overset{ \nu\ n}{\longrightarrow}}\emptyset,\ \
\emptyset{\overset{\mu_l }{\longrightarrow}}{\cal A},\ ,\ \mu_l=\mu l,\ l=0,1,\ \ 
{\cal O}^A_0+{\cal A} \underset{\kappa }{\overset{  g(n)}{\longleftrightarrow}}{\cal O}^A_1,$$
where the symbol ${\cal A}$ stands for  activator proteins, and
${\cal  O}^A_l$, $l=0,1$ denotes the state of the
promoter related to the activator, and by the chemical reactions related to therapeutic proteins as given by
$$
{\cal  O}^T_0+{\cal A} \underset{\hat\kappa }{\overset{  h(n)}{\longleftrightarrow}}{\cal \hat O}^T_1,
{\cal X}{\overset{\hat\nu x}{\longrightarrow}}\emptyset,\ \
\emptyset{\overset{\hat\mu_l }{\longrightarrow}}{\cal X},\ \hat\mu_l=\hat\mu l,\ l=0,1,
$$
where ${\cal A}$ denotes activator proteins,  and
${\cal  O}^T_l$
is defined in a similar way for the promoter of the
therapeutic gene and ${\cal X}$ symbolizes therapeutic proteins. 
 
\subsection{ Equilibrium equations\label{s.equilibrium}}

The modeling of the time evolution of the number of 
molecules involved in the regulatory network
is obtained by assuming that extrinsic noise,
here the random fluctuations of the number of
repressor and doxycycline molecules  attains
a chemical equilibrium. We first describe mathematically
the effect of this extrinsic noise on the promoters
associated to the activator and therapeutic genes.
We follow Section 28 of \cite{Dill}. Consider a multiple binding
of a ligand X with $1\le i\le k$ different binding sites on a  polymer P,
$$
P + iX\longrightarrow PX_i,\ i=1\cdots k,
$$
with equilibrium constants
$$K_i=\frac{[PX_i]}{[P][X]^i}.$$
The {\it binding polynomial} is defined by
$$Q(X)=1+\sum_{i=1}^k K_i X^i,$$
where in the sequel $X$ will denote the number of ligand molecules. The
proportion of P molecules that are in the $i$-th liganded state is
$\displaystyle \frac{[PX_i]}{[P]\ Q}$, and the average number of bound sites is
$$M(X)=\frac{{\rm d}\ln(Q)}{{\rm d}\ln(X)}=\frac{\sum_{i=0}^k i K_i X^i}{Q(X)}.$$
\begin{example}\label{independent}
If the $k$ binding sites are independent, there is no cooperativity, and one has
$$Q(X)=(1+KX)^k,$$
with
$$M(X)=\frac{k K X}{1+K X}.$$
\end{example}
\begin{example}\label{Hill}
If a P molecule binds to exactly $k$ ligands molecules at a time, one gets
the Hill model $k X+P\longrightarrow PX_k$, with equilibrium constant $K$,
and
$$Q(X)=1+K X^k,$$
$$M(X)=\frac{k K X^k}{1+K X^k}.$$
\end{example}

\subsection{Transgene expression}
      
  \centerline{\bf Reaction of  TetR repressor and doxycycline}
\medskip

The reaction between the doxycycline (Dox)
and the repressor (R) is described as $R+Dox\longrightarrow RD$ with some
forward rate, and $RD\longrightarrow R+D$ with some backward rate; Considering
equilibrium of constant $K_{RD}$, we can write
$
K_{RD}=[RD]/([R][Dox])$,
where $[Dox]$ gives the number of molecules of doxycycline. The free proportion
of (R) molecules, i.e. not involved in the RD complex, can, when considered as
ligand, bind to the $k_r$ sites of the TetR operators
(the binding sites where repressor molecules can bind, see e.g. \cite{Mermod}), inhibating thus
both the transactivator and the synthesis of the transgene product. We next estimate
 the average fraction
$F(R,[Dox])$
of  sites free of repressor. Let $k_r$ denote the number of
sites where repressors molecules can bind. Using a Hill model
of cooperativity (see Example \ref{Hill}), one gets
that
the average number of bound sites is then given by
$$M_r([R])=\frac{k_r K_r [R]^{k_r}}{1+K_r [R]^{k_r}},$$
Then,
\begin{equation*}
F(R,[Dox])=\frac{k_r -M_r([R])}{k_r}=\frac{1}{1+K_r [R]^{k_r}}.
\end{equation*}
The total number of repressor, denoted by $R_{tot}$, is such that
$$[R_{tot}]\approx [R]+K_{RD}[R][Dox])=[R](1+K_{RD}[Dox]),$$
when we neglect the amount of repressor involved in the $k_r$ binding
sites. Set $[R_{tot}]=R_{\max}$. Then
\begin{equation*}\label{newfraction}
F(R,[Dox])=\frac{(1+K_{RD}[Dox])^{k_r}}{(1+K_{RD}[Dox])^{k_r}+K_r R_{\max}^{k_r}}.
\end{equation*}
\bigskip

 \centerline{\bf Transactivator }
 \bigskip
 The transactivator is repressed by the bound repressors,
 and activated by the positive feedback loop;
 The above considerations suggest a stochastic model
 of transactivation  with
 \begin{equation*}\label{FeedbackDox}
 g(d)= V F(R,[Dox]) d^a+g(0),
 \end{equation*}
 where $a$ denotes the number of binding sites on the
 activator, $d$ denotes the number of transactivator dimers, and where $V$ is a parameter.
 $g(0)>0$ is introduced here to model basal activity for the off to on transitions.
 \medskip

\subsection{The regulatory network at quasi-equilibrium\label{s.networkquasi}}
\bigskip

 We assume that
the promoter switch from the off to on state
at rate $$g(D^\eps(t))=V F(R,[Dox])(D^\eps(t))^a+g(0),$$
where $D^\eps(t)$ is the number of transactivator
dimers present at time $t$, $a$ is the number of
activator binding sites and $V$ is a parameter.
This models the positive feedback loop.
We suppose that degradation occurs at a rate
proportional to the number of monomers
$N^\eps(t)-2D^\eps(t)$, with constant of proportion
$\nu$. The transactivator process is given
by the triplet $(N^\eps(t),Y^\eps(t),D^\eps(t))$,
where we assume fast dimerization, as given
in the preceeding paragraph.
The time evolution of the network is described
 by the random process
$$\eta^\eps(t)=(N^\eps(t),Y^\eps(t),D^\eps(t),X^\eps(t),Z^\eps(t)),$$
where $X^\eps(t)$ denotes the number of therapeutic
proteins $({\cal X})$ present in the cell at time $t$, and
where $Z^\eps(t)=0,1$ denotes the state of its
associated promoter (off/on). These chemical reactions
are described schematically as

$$
{\cal M}{\overset{\nu m }{\longrightarrow}}\emptyset,\ \
\emptyset{\overset{\mu_l }{\longrightarrow}}{\cal M},\ l=0,1,\ \ 
{\cal O}_0^A+{\cal D} \underset{\kappa }{\overset{ g(d)}{\longleftrightarrow}}{\cal O}_1^A,
{\cal M}+{\cal M}\underset{\beta_-^{\varepsilon} }{\overset{\beta_+^{\varepsilon}}{\longleftrightarrow}}{\cal D},
$$
$$
{\cal  O}_0^T+{\cal D} \underset{\hat\kappa }{\overset{ h(d)}{\longleftrightarrow}}{\cal  O}_1^T,
{\cal X}{\overset{\hat\nu x}{\longrightarrow}}\emptyset,\ \
\emptyset{\overset{\hat\mu_l }{\longrightarrow}}{\cal X},\ l=0,1,
$$
 where  ${\cal  O}_l^T$, $l=0,1$ accounts
for the state of the promoter related to the therapeutic gene, 
${\cal M}$ denotes activator proteins (monomers), $\mu_l=\mu l$, $l=0,1$,
${\cal X}$ denotes therapeutic proteins and $\hat\mu_l=\hat\mu l$, $l=0,1$.
We again assume a quasi-equilibrium with fast
dimerization, to get the limiting process 
$$\eta(t)=(N(t),Y(t),X(t),Z(t)),$$
associated with the set of coupled chemical reactions
$$
{\cal A}{\overset{\nu n }{\longrightarrow}}\emptyset,\ \
\emptyset{\overset{\mu_l }{\longrightarrow}}{\cal A},\ l=0,1,\ \ 
{\cal O}_0^A+{\cal A} \underset{\kappa }{\overset{ g(n)}{\longleftrightarrow}}{\cal O}_1^A,
$$
$$
{\cal  O}_0^T+{\cal A} \underset{\hat\kappa }{\overset{ h(n)}{\longleftrightarrow}}{\cal  O}_1^T,
{\cal X}{\overset{\hat\nu x}{\longrightarrow}}\emptyset,\ \
\emptyset{\overset{\hat\mu_l }{\longrightarrow}}{\cal X},\ l=0,1,
$$
with quasi-equilibrium transition rates given by (see Section \ref{s.two})
\begin{align*}\nu(n)&=\nu\E_{\mu^n}(n-2d),\\ g(n)&=V F(R,[Dox])\E_{\mu^n}(d^a)+g(0), \\
 h(n)&=\hat V F(R,[Dox])\E_{\mu^n}(d^a)+h(0).\end{align*}

\subsection{A semi-stochastic mean field model\label{s.meanfieldnet}}
\begin{figure}
  \centering
  \includegraphics[width=\textwidth]{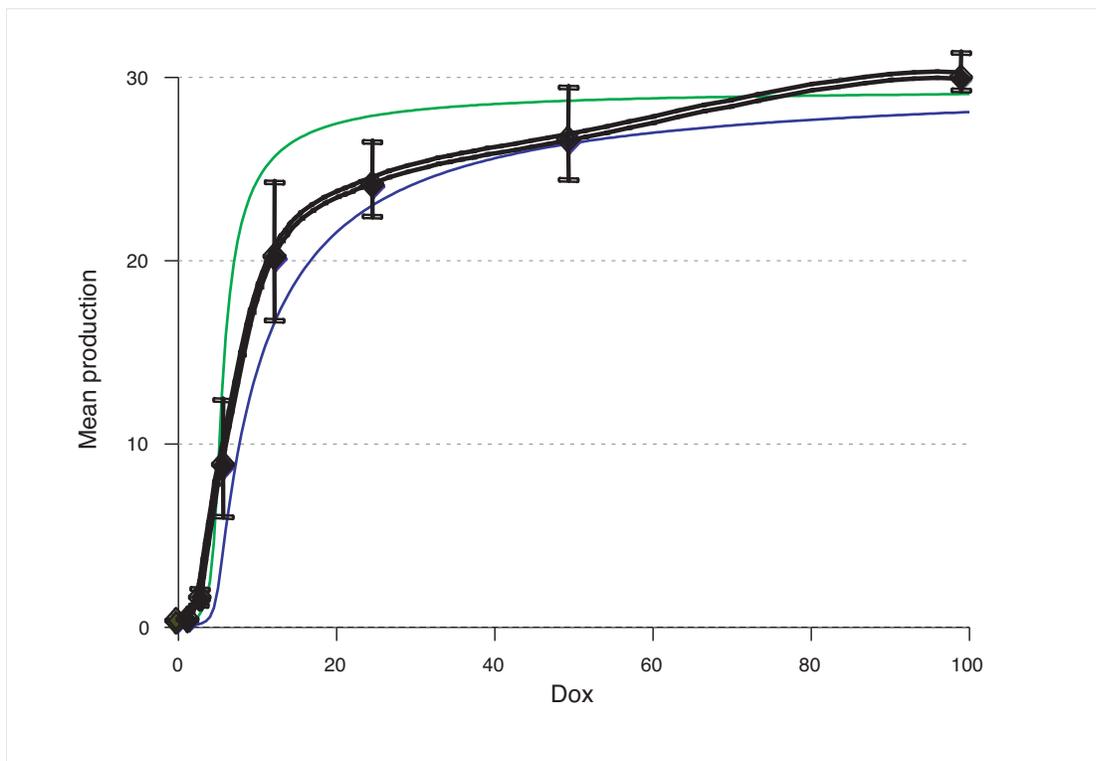}
  \caption{ 
  Assay of the regulation of EGFP expression in living cells. The experimental
  curve obtained in \cite{Fournier}, (in black) giving the average expression
  of therapeutic proteins as function of the number [Dox] of doxycycline molecules
  is compared to the curves  obtained from the mean-field model, where the green and blue curves
  provide the mean expression levels of activator and therapeutic proteins, respectively.
  }
  \label{fig5}
\end{figure}

We consider the time evolution of the network in a semi-stochastic
version by supposing that the rates $g(N(t))$ and
$h(N(t))$ are replaced by  $c(t)=g(\E(N(t-\theta)))$ and
$\hat c(t)=h(\E(N(t-\hat\theta)))$. The method is similar to what is presented 
in Section \ref{s.meanfield}. Consider the family of transition kernels
$L_{c,\hat c}=(q_{c\hat c}((n,y,x,z)(n',y',x',z'))$ given by
$$q_{c\hat c}((n,y,x,z)(n+1,y,x,z))=y\mu,\ \ 
q_{c\hat c}((n,y,x,z)(n-1,y,x,z))=\nu\ n,$$
$$q_{c\hat c}((n,y,x,z)(n,1-y,x,z))=\kappa y +(1-y)c,\ \ 
q_{c\hat c}((n,y,x,z)(n,y,x+1,z))=z \hat\mu,$$
$$q_{c\hat c}((n,y,x,z)(n,y,x-1,z))=\hat\nu x,\ \ 
q_{c\hat c}((n,y,x,z)(n,y,x,1-z))=\hat\kappa z +(1-z)\hat c.$$
The nice feature of this kernel is that its steady state distribution
is the product $\pi^{\kappa,c}\otimes\pi^{\hat\kappa,\hat c}$ of the stationary distributions
associated with the self regulated genes given by the two sets of chemical reactions
$$
{\cal A}{\overset{\nu n }{\longrightarrow}}\emptyset,\ \
\emptyset{\overset{\mu_l }{\longrightarrow}}{\cal A},\ l=0,1,\ \ \ 
{\cal O}_0 \underset{\kappa }{\overset{ c}{\longleftrightarrow}}{\cal O}_1,
$$
$$
{\cal X}{\overset{\nu x }{\longrightarrow}}\emptyset,\ \
\emptyset{\overset{\hat\mu_l }{\longrightarrow}}{\cal X},\ l=0,1,\ \ \ 
{\cal \hat O}_0 \underset{\hat\kappa }{\overset{ \hat c}{\longleftrightarrow}}{\cal \hat O}_1,
$$
Both measures can be computed efficiently by using either the method of transfer matrices
or the exact analytical expressions obtained through generating functions.
Coming back to the time evolution of the network under a mean field model, the method
is similar to that given in Section \ref{s.meanfield} and consists in two basic steps:
\begin{itemize}
\item{} Find the limiting values $c(\infty)=\lim_{t\to\infty}c(t)$ and $\hat c(\infty)=\lim_{t\to\infty}\hat c(t)$,
when they exist,
\item{} compute the steady state distribution $\pi^{\kappa,c(\infty)}\otimes\pi^{\hat\kappa,\hat c(\infty)}$, and
              the related means and variances.
\end{itemize}
Let $(N(\infty),Y(\infty),X(\infty),Z(\infty))$ be distributed according to the steady state distribution.
Proceeding as in Example \ref{Constant}, the coefficient of variation related to the activator satisfies
$$CV^2_N=\frac{{\rm Var}(N(\infty))}{\E(N(\infty))^2}=
\frac{1}{\E(N(\infty))}+\frac{\tau_2}{\tau_1+\tau_2}\frac{{\rm Var}(Y(\infty))}{\E(Y(\infty))^2},$$
where 
$$\E(N(\infty))=\frac{\mu}{\nu}\frac{c(\infty)}{c(\infty)+\kappa},\ 
\tau_1=\frac{1}{\nu},\ \hbox{ and } \tau_2=\frac{1}{c(\infty)+\kappa}.$$
Similarly the CV of the transgene product is such that
$$CV^2_X=\frac{{\rm Var}(X(\infty))}{\E(X(\infty))^2}=
\frac{1}{\E(X(\infty))}+\frac{\hat\tau_2}{\hat\tau_1+\hat\tau_2}\frac{{\rm Var}(Z(\infty))}{\E(Z(\infty))^2},$$
where 
$$\E(X(\infty))=\frac{\hat\mu}{\hat\nu}\frac{\hat c(\infty)}{\hat c(\infty)+\hat\kappa},$$
$$\hat\tau_1=\frac{1}{\hat\nu},\ \hbox{ and } \hat\tau_2=\frac{1}{\hat c(\infty)+\hat\kappa}.$$
In what follows, we consider $c(\infty)$ and $\hat c(\infty)$.

\subsubsection{$c(\infty)$ and $\hat c(\infty)$ for linear feedbacks\label{s.semistoch.lin}}

In the linear case, $c(t)=g(\E(N(t-\theta)))=g_0+g_1 \E(N(t-\theta))$
and $\hat c(t)=h(\E(N(t-\theta)))=h_0+h_1 \E(N(t-\hat\theta))$. We thus consider
the averages
$$E(t)=\E(N(t)),\ \ G(t)=\E(Y(t))\hbox{ and }
\hat E(t)=\E(X(t)),\ \ \hat G(t)=\E(Z(t)),$$
which satisfy the system of delayed differential equations

$$\frac{dG(t)}{dt}=(g_0+g_1 E(t-\theta))(1-G(t))-\kappa G(t),\ \ \frac{dE(t)}{dt}=\mu G(t)-\nu E(t),$$
$$\frac{d\hat G(t)}{dt}=(h_0+h_1  E(t-\hat\theta))(1-\hat G(t))-\kappa \hat G(t),\ \ 
\frac{d\hat E(t)}{dt}=\hat\mu \hat G(t)-\hat\nu \hat E(t).$$

The results of Section \ref{s.linear} yield that
the above delayed differential system has a globally asymptotically
stable equilibrium point 
$(E(\infty),G(\infty),\hat E(\infty),\hat G(\infty))$, with
$$E(\infty)=\frac{\mu}{\nu}G(\infty),\ \ 
(g_0+g_1\frac{\mu}{\nu}G(\infty))(1-G(\infty))=\kappa G(\infty),$$
$$(h_0+h_1 \frac{\mu}{\nu}G(\infty))(1-\hat G(\infty))=\ \hat\kappa \hat G(\infty) \hbox{ and }\hat E(\infty)=\frac{\hat\mu}{\hat\nu}\hat G(\infty).$$
Finally
$$c(\infty)=g_0+g_1 E(\infty)\hbox{ and }\hat c(\infty)=h_0+h_1 \hat E(\infty).$$
\section{Conclusion and discussion\label{conclusion}}
In this work, we considered a class of self-regulated genes which are
the building blocks of most of the existing gene networks. We provided efficient numerical
algorithms for computing the  steady state distribution of the number of produced
proteins. These results permit to handle more complex situations, and to understand 
the effect of positive  or negative feedbacks in the network's dynamics. They might also
be useful in reverse engineering problems when infering for example  the parameters defining
chemical reactions. Next, we considered in Sections 3 and 5 mean field models with time delays which are of special interest in
synthetic biology or in biotechnology, where small engineered regulatory networks are
inserted at random in host genomes.  Mathematical results in this setting are very 
scarce, and it is known that such systems can exhibit oscillations (see e.g. \cite{Delayed}
or \cite{MeanField}). Section 3 provides convergence results
for mean field models with time delays, which might open
ways for handling more complex gene networks.
Experimental results performed in
living cells were in good concordance with our predictions. This shows that 
such models can provide relevant informations concerning complex systems, and that 
mathematical models can be efficiently used for the design of new regulatory gene
networks in synthetic biology or in biotechnology.

\section{Appendix\label{s.appendix}}

\subsection{Fluctuation Lemma}\label{FlucLemma2}
The following result is a slight modification of Lemma $4.2$ in \cite{Hirsch}:
\begin{lemma}\label{FlucLemma}
Let $f:\R_+\rightarrow \R$ be bounded and differentiable, $\dot f$ denoting its derivative. There exist increasing sequences $t_n\uparrow+\infty$ and $s_n\uparrow+\infty$ , such that $$f(t_n)\rightarrow\overline f, \ \dot f(t_n)\rightarrow0,
\text{ and }
f(s_n)\rightarrow \underline f, \ \dot f(s_n)\rightarrow0$$ as $n\rightarrow+\infty$, where for a function $f$   we denote $$\overline f:=\limsup_{t\rightarrow+\infty}f(t),\ \underline f:=\liminf_{t\rightarrow+\infty}f(t).$$
\end{lemma}

\subsection{Convergence of time-nonhomogeneous Markov Chains\label{nonhomogeneous2}}

We consider a nonhomogeneous Markov chain $X(t)$ taking values in $\N$,
of instantaneous transition matrix $Q_t=(q_t(i,j))_{i,j\in\N}$.
The following Theorem is proved in \cite{Abramov}.
\begin{theorem}\label{nonhomogeneous}
Assume that we can find nonnegative constants $q(i,j)$ such that
$$\sum_{j\ne i}q(i,j)<+\infty,\ \ \int_0^\infty(\sqrt{q_t(i,j)}-\sqrt{q(i,j)})^2 {\rm d}t<+\infty,$$
and
$$\int_{0\le s\le t,\ q(i,j)>0}q_s(i,j){\rm d}s = \int_0^t q_s(i,j) {\rm d}s.$$
Let $Q_0=(q(i,j))_{i,j\in\N}$, and let $X^0(t)$ be the related $\N$-valued Markov chain.
Suppose that $Q_0$ is ergodic, that is that there is a unique probability measure $\pi$
such that $\pi Q_0=0$ and 
$$\lim_{t\to\infty}P(X^0(t)=j\vert X^0(s)=i)=\pi_j,\ \ \forall s,\ i,\ j.$$
Then
$$\lim_{t\to\infty}P(X(t)=j\vert X(s)=i)=\pi_j,\ \ \forall s,\ i,\ j.$$
\end{theorem}

\subsection{Zeros of an exponential polynomial\label{expopoly2}}

We consider the exponential polynomial $H(z)=(z^2+pz+q)e^z+r$, where $p$ is real and positive, $q$ is real and nonnegative, and $r$ is real. 
The following Theorem is proved in \cite{Bellman}, p. 449.
\begin{theorem}\label{expopoly}
Denote by $a_k$ $(k\geq 0)$ the sole root of the equation $\cot(a)=(a^2-q)/p$ which lies on the interval $(k\pi,k\pi+\pi)$. We define the number $w$ as follows:
\begin{enumerate}
\item if $r\geq 0$ and $p^2\geq 2q$, $w=1$;
\item if $r\geq 0$ and $p^2< 2q$, $w$ is the odd $k$ for which $a_k$ lies closest to $\sqrt{q-p^2/2}$;
\item if $r< 0$ and $p^2\geq 2q$, $w=2$;
\item if $r< 0$ and $p^2< 2q$, $w$ is the even $k$ for which $a_k$ lies closest to $\sqrt{q-p^2/2}$.
\end{enumerate}
Then, a necessary and sufficient condition that all roots of $H(z)=0$ lie to the left of the imaginary axis is that
\begin{enumerate}
\item $r\geq 0$ and $r\sin(a_w)/(pa_w)<1$ or
\item $-q<r< 0$ and  $r\sin(a_w)/(pa_w)<1$.
\end{enumerate}
\end{theorem}
 
\eject


 \newpage

\end{document}